\documentclass[aps,pre,superscriptaddress,twocolumn,longbibliography,groupedaddress]{revtex4-2}

\usepackage{float}
\usepackage{graphicx}
\usepackage{caption}
\usepackage{subcaption}
\usepackage{amsmath,amsthm,amssymb}
\usepackage{amsfonts}
\usepackage{dcolumn}
\usepackage{bm}
\usepackage{epstopdf}
\usepackage{algorithm}
\usepackage{algpseudocode}
\usepackage[colorlinks]{hyperref}

\usepackage{epsfig}
\usepackage{mathrsfs}
\usepackage{multirow}
\usepackage{tabularx}
\usepackage[all]{xy}
\usepackage{pbox}
\usepackage{verbatim}
\usepackage{mathtools}
\usepackage{tikz,pgfplots}
\usepackage{xcolor}
\usepackage{xfrac}
\usepackage{cleveref}
\usepackage{booktabs}
\usepackage{physics}
\usepackage{siunitx}
\usepackage{url}
\usepackage{hyperref}
\usepackage{afterpage}
\usepackage[normalem]{ulem}

\DeclareMathOperator{\ord}{ord}

\DeclareMathOperator{\lcm}{lcm}

\newcommand{\opcat}[1]{{#1}_{\mathrm{TPM}}}
\newcommand{\opcatq}[1]{{#1}_{\mathrm{QTPM}}}

\newcommand{\er}[1]{Eq.\eqref{#1}}
\newcommand{\ers}[2]{Eqs.(\ref{#1}-\ref{#2})}

\newcommand{\singlefigure}{0.45\textwidth}

\newlength\figureheight 
\newlength\figurewidth 

\hypersetup{
  colorlinks,
  citecolor=violet,
  linkcolor=violet,
  urlcolor=blue}

\begin{document}

\title{Boundary conditions dependence of the phase transition in the quantum Newman-Moore model}

\author{Konstantinos Sfairopoulos}
\email{ksfairopoulos@gmail.com}
\affiliation{School of Physics and Astronomy, University of Nottingham, Nottingham, NG7 2RD, UK}
\affiliation{Centre for the Mathematics and Theoretical Physics of Quantum Non-Equilibrium Systems,
University of Nottingham, Nottingham, NG7 2RD, UK}
\author{Luke Causer}
\affiliation{School of Physics and Astronomy, University of Nottingham, Nottingham, NG7 2RD, UK}
\affiliation{Centre for the Mathematics and Theoretical Physics of Quantum Non-Equilibrium Systems,
University of Nottingham, Nottingham, NG7 2RD, UK}
\author{Jamie F. Mair}
\affiliation{School of Physics and Astronomy, University of Nottingham, Nottingham, NG7 2RD, UK}
\affiliation{Centre for the Mathematics and Theoretical Physics of Quantum Non-Equilibrium Systems,
University of Nottingham, Nottingham, NG7 2RD, UK}
\author{Juan P. Garrahan}
\affiliation{School of Physics and Astronomy, University of Nottingham, Nottingham, NG7 2RD, UK}
\affiliation{Centre for the Mathematics and Theoretical Physics of Quantum Non-Equilibrium Systems,
University of Nottingham, Nottingham, NG7 2RD, UK}

\begin{abstract}
We study the triangular plaquette model (TPM), also known as the Newman-Moore model, in the presence of a transverse magnetic field on a lattice with periodic boundaries in both spatial dimensions. We consider specifically the approach to the ground state phase transition of this quantum TPM (QTPM), or quantum Newman-Moore model, as a function of the system size and type of boundary conditions. Using methods based on cellular automata, we obtain a characterisation of the minimum energy configurations of the TPM for numerically accessible tori sizes. For the QTPM, we use these cycle patterns to obtain the symmetries of the model which, we argue, indicate the nature of its quantum phase transition: we always identify it as a first-order phase transition, with the addition of spontaneous symmetry breaking for system sizes which have degenerate classical ground states. For sizes accessible to numerics, we corroborate our findings with exact diagonalization, matrix product states and quantum Monte Carlo simulations. 
\end{abstract}

\maketitle

\section{Introduction}

In this paper, we study the ground state phase transition of the quantum Newman-Moore model, or quantum triangular plaquette model. The classical triangular plaquette model (TPM), introduced by Newman and Moore \cite{1999_Newman_Moore}, is a model of Ising spins interacting in triplets in (half of) the plaquettes of a triangular lattice. Despite the absence of quenched disorder and its trivial static properties, the model has rich glassy dynamics \cite{1999_Newman_Moore,Garrahan2000,2002_Garrahan}. The TPM is an important model as it realises at low temperatures the paradigm of slow (super-Arrhenius) relaxation due to effective kinetic constraints in an interacting system. This phenomenon is central to the dynamic facilitation picture of the glass transition \cite{2010_Chandler,2019_Speck,2021_Hasyim}. The physics of the TPM can also be generalised to  three dimensions, for example in the five-spin interaction square-pyramid model \cite{2016_Jack}, or maintaining the triangular interactions in the models of Ref.~\cite{2022_Biswas}. A three-dimensional generalisation of the TPM with non-commuting terms \cite{2005_Chamon} actually  started what is now the field of fractons \cite{2018_Devakul_Sondhi,2019_Nandkishore_Fractons,2020_Pretko,2022_McGreevy}.

The simplest way to transform the TPM into a quantum model is by adding a transverse field term to the classical Hamiltonian. Such quantum TPM (QTPM), or quantum Newman-Moore model was considered in the context of fractons in Refs.~\cite{2013_Yoshida,2014_Yoshida,2019_Devakul}. Numerics in Ref.~\cite{2014_Yoshida} suggested that the ground state undergoes a first-order transition. A related work studying the large deviations of plaquette observables in the stochastic dynamics of independent spins \cite{2020_Vasiloiu} also found numerical evidence for a first-order transition at the self-dual point of the model. 

In contrast, the results from Ref.~\cite{2021_Zhou_Pollmann} indicated that the transition is continuous, with a particular form of fractal symmetry breaking. The classical TPM and its connection with fractals and topological order was also drawn in Ref.~\cite{2013_Yoshida}. Here we aim to resolve these discrepancies, primarily regarding the interpretation of its quantum phase transition, by exploiting a general connection between $D$-dimensional cellular automata (CA) \cite{1983_Wolfram} and the ground states of $(D+1)$-dimensional classical spin models \cite{2023_Sfairopoulos_2}. By using this method in the specific case of the QTPM with periodic boundary conditions, we are able to characterize the approach to its quantum phase transition in the large size limit. Our key observation is that the nature of the transition depends on the specific lattice dimensions, and this is manifested in the finite size scaling. 

For the TPM, the relevant CA is Rule 60 \cite{1983_Wolfram} and not Rule 90 that might be assumed from comment [18] in Ref.~\cite{1999_Newman_Moore}. For system sizes where one dimension is a power of two, Rule 60 has a single fixed point \cite{2005_Calkin}, implying a single energy minimum for the classical TPM. In such cases, we verify that the quantum phase transition in the QTPM is of first-order (that is, a sequence of such system sizes tends to a first-order transition in the large size limit). This also holds for other sizes for which Rule 60 has no non-trivial attractors. However, for certain sizes there can be periodic orbits on top of the fixed point for the CA, giving rise to classical ground state degeneracies in the TPM. For the quantum model, this translates to a mixed order quantum phase transition. We provide evidence for this scenario by means of numerical simulations, namely, for small sizes using Exact Diagonalisation, and for large sizes using both Matrix Product State approximations of the ground state, and continuous-time Quantum Monte Carlo \cite{1996_Beard, 2008_Krzakala, 2012_Mora, 2023_Causer}. 

The paper is organised as follows. In Sec.~\ref{TPMclassical_quantum}, we review the classical and quantum TPM. In Sec.~\ref{CA_section}, we provide the necessary background on CA and the connection to the ground states of the classical TPM. In Sec.~\ref{PT}, we discuss the ground state phase transition of the QTPM in terms of the symmetries that follow from the properties of the associated CA, and support our predictions with numerical simulations. In Sec.~\ref{conclusions} we give our conclusions. In Appendix~\ref{appendixA} we provide further details, including the case of the QTPM with open boundaries, while in Appendix~\ref{appendixB} we discuss the scaling of the energy gap.
\section{Triangular Plaquette Model, classical and quantum}{\label{TPMclassical_quantum}}

\subsection{Classical}
\begin{figure}
    \centering
    \includegraphics[width=\singlefigure]{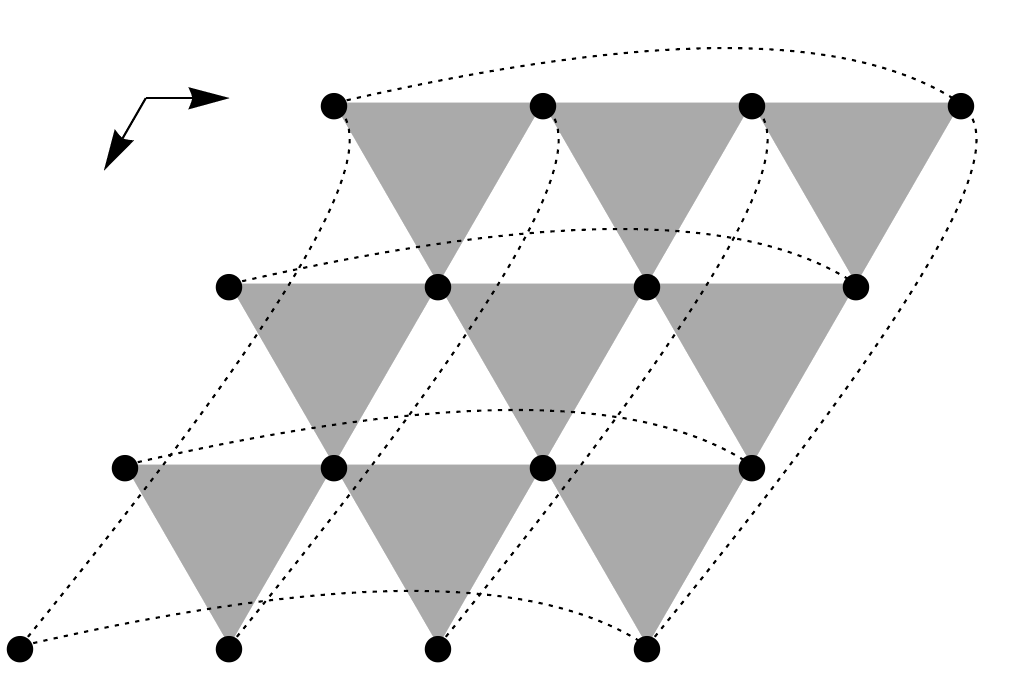}
    \caption{\label{fig:TPM_lattice} 
    Triangular plaquette model. The shaded triangles indicate the interacting triplets, \er{TPMclassical}. Dotted lines indicate spins that are identified by periodic boundary conditions on an $N=4 \times 4$ lattice.
    }
\end{figure}

The triangular plaquette model \cite{1999_Newman_Moore,Garrahan2000,2002_Garrahan} is a model of Ising spins $s_i = \pm 1$ on the sites $i = 1, \ldots, N$ of a triangular lattice, with cubic interactions between the spins on the vertices of downward-pointing triangles of the lattice in Fig.~\ref{fig:TPM_lattice}. The Hamiltonian of this classical model reads
\begin{equation}{\label{TPMclassical}}
    \opcat{E} =  - J \sum_{\{i, j, k\} \in \triangledown} s_i s_j s_k .
\end{equation} 
In what follows, it will be convenient to consider the equivalent model on a square lattice of size $N = L \times M$, with classical Hamiltonian, 
\begin{equation}
    \opcat{E} =  - J \sum_{x,y = 1}^{L,M} s_{x,y} s_{x+1,y} s_{x+1, y+1},
    \label{Hsq}
\end{equation}
where we assume periodic boundary conditions (PBC) in both directions by identifying 
\begin{align*}
    x + L &= x \mod{L} \\
    y + M &= y \mod{M}.
\end{align*}
In \er{Hsq}, we label the spins by $s_{x,y}$ at site with coordinates $(x,y)$ in the square lattice.

The classical TPM has been predominantly studied in the context of the glass transition. For lattices with at least one dimension being a power of two, the energy \er{TPMclassical} reduces to that of the non-interacting plaquette variables \cite{1999_Newman_Moore,Garrahan2000,2002_Garrahan},
\begin{equation}
    \opcat{E} = - J \sum_{\triangledown} d_\triangledown,
    \label{plaq}
\end{equation}
where $d_\triangledown = s_i s_j s_k$ with $\{i, j, k\} \in \triangledown$ for every downward-pointing triangle in the lattice. When at least one dimension is a power of two, the relation between plaquettes and spin variables is one-to-one, exactly proving the above. The thermodynamics of the TPM is therefore one of free binary excitations and, as such, it is essentially trivial.

In contrast to the statics, the single spin-flip dynamics of the TPM is highly non-trivial, as flipping one spin changes three adjacent plaquettes. This implies that at low temperatures, where excited plaquettes are suppressed, cf.~\er{plaq}, the dynamics has effective kinetic constraints \cite{Garrahan2000}. These dynamical constraints lead to an activated relaxation similar to that of the East model \cite{2002_Ritort_Sollich_review}, with relaxation times growing as the exponential of the inverse temperature squared \cite{Garrahan2000} (a super-Arrhenius form known as the ``parabolic law'' \cite{2009_Elmatad}). Similar glassy behaviour is seen  in generalisations of the TPM with odd plaquette interactions \cite{2015_Garrahan_Turner,2022_Biswas}. The TPM has also been considered in the presence of a (longitudinal) magnetic field \cite{2010_Sasa} and in the related case of coupled replicas \cite{2014_Garrahan}, and the TPM with open boundary conditions was studied in Ref.~\cite{2012_Yamaguchi} through partial
trace methods. The classical TPM was studied in the context of topological order in fractal models in Ref.~\cite{2013_Yoshida}. Autoregressive neural networks were applied to the classical TPM with limited success in \cite{2022_RMelko_TPM}. 

A diluted ferromagnetic model with $p$-spin interactions with $p=3$ on hypergraphs was studied in Ref.~\cite{2001_Franz} using replica techniques. In contrast, the TPM is defined on a 2D square lattice with interactions only on downward-pointing triangles. Both models were found to be glassy, although the origin of their glassiness was attributed to different mechanisms \cite{2001_Franz,1999_Newman_Moore,2002_Garrahan}. In the computer science terminology $p$-spin models correspond to constrained satisfaction problems,  specifically to $p$-XORSAT models \cite{montanari_book}. The TPM, thus, corresponds to a specific instance of the unfrustrated 3-regular 3-XORSAT \cite{2001_Ricci-Tersenghi,2005_Castellani}, since its spin participates in exactly three plaquette interactions.

\subsection{Quantum: TPM in a transverse field}

Taking \er{TPMclassical} and adding a transverse field, we obtain the Hamiltonian of the quantum TPM (QTPM), 
\begin{equation}{\label{QTPM}}
    \opcatq{H} =  - J \sum_{\{i, j, k\} \in \triangledown} Z_i Z_j Z_k - h \sum_i X_i,
\end{equation}
where $Z_i$ and $X_i$ represent Pauli operators acting non-trivially on site $i$. This quantum model was studied in Refs.~\cite{2014_Yoshida,2019_Devakul} and its connections to models of fractons were investigated.

The Hamiltonian \eqref{QTPM} is expected to have a quantum phase transition at the self-dual point $J=h$ \cite{2020_Vasiloiu}.
Dualities emerge in various physical situations \cite{2011_Cobanera_BondAlgebraic}. In this specific case, the duality embodies the relation between the classical and the quantum paramagnetic phases of the same model. In standard statistical mechanics, they were first identified in Ref.~\cite{1941_Kramers}, relating the low-temperature phase of the classical 2D Ising model with the high temperature phase of the same model (for reviews see~\cite{1979_Kogut,1980_Savit}). The location of the phase transition (if any) of the given model consists of the main prediction of the duality arguments. For the duality in the case of the TPM see Refs.~\cite{2020_Vasiloiu,2021_Zhou_Pollmann}. 

Numerical results from Refs.~\cite{2014_Yoshida,2020_Vasiloiu} suggested the quantum phase transition of the TPM to be of first-order. 
    This implies a discontinuity of the first derivative of the ground state energy 
    as an indicator of the phase transformation.   
Ref.~\cite{2020_Vasiloiu} arrived to these conclusions by using trajectory sampling in systems with linear size a power of two and PBC. In contrast, Ref.~\cite{2021_Zhou_Pollmann} found evidence for a continuous phase transition with fractal symmetry breaking using stochastic series expansion methods, also with periodic boundary conditions but not restricted to power of two sizes.
The study of QTPM was connected to Rydberg atoms in Ref.~\cite{2022_Xu_Myerson-Jain}. Further studies on the QTPM and its generalisations from the viewpoint of fracton field theory were presented in Refs.~\cite{2022_Xu_Myerson_2,2022_Xu_Vijay_Myerson}. 

In the context of quantum adiabatic evolution and quantum annealing \cite{2000_Farhi}, 3-XORSAT instances with a transverse field similar to the TPM have been shown to possess a first-order quantum phase transition \cite{2012_Farhi_Gosset,2019_Kourtis_Chamon} and classical glassy behaviour \cite{2019_Kourtis_Chamon}. Both hinder the efficiency of (classical or quantum) annealing. Interestingly, the transition point found in Ref.~\cite{2012_Farhi_Gosset} 
coincides with that we find here for the QTPM. 
We also note that the study of instances of the 3-XORSAT model with multiple degeneracies was associated to a continuous quantum phase transition in Ref.~\cite{2021_Medina} based on the related energy spectrum.

\section{Cellular Automata and ground states of the classical TPM}{\label{CA_section}}

\subsection{General aspects of CA}

Cellular automata (CA) consist of a $D$-dimensional array of sites evolving under discrete time and synchronous dynamics. 
Under this evolution, the state is updated by a deterministic rule which is local in time (although generalisations exist) \cite{1983_Wolfram,1984_Wolfram,1993_Stevens,1999_Stevens}. As a result, CA dynamics gives rise to, often rich, $(D+1)$-dimensional structures. In what follows, we consider $D=1$, linear CA with deterministic local transition rules with sites taking values from the finite field $\mathbb{F}_2$.

\begin{figure}
    \centering
    \begin{subfigure}[b]{0.45\textwidth}
        \includegraphics[width=1\linewidth]{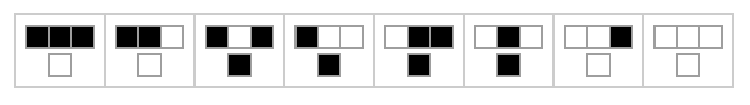}
        \caption{}
     \end{subfigure}
     \begin{subfigure}[b]{0.45\textwidth}
        \includegraphics[width=1\linewidth]{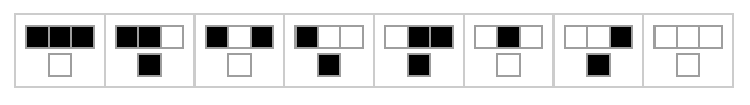}
        \caption{}
     \end{subfigure}
     \caption[Two Rule Plots]{(a) Rule 60 and (b) Rule 90 evolution rules.}
     \label{fig:Ruleplots}
\end{figure}

Discrete cellular automata are fully specified by an initial configuration of $L$ sites and a \textit{local transition rule}. The local update rule, $f$, determines the configuration of every site at each timestep using the local neighbourhood of size $2r +1$, given its radius $r$. For $D=1$, elementary CA rules are defined by $r=1$ \cite{1983_Wolfram,1984_Wolfram}. For a neighbourhood of three sites, $r=1$, there are thus 8 possible configurations giving rise to 256 possible choices for update rules. Figure \ref{fig:Ruleplots}(a) shows Rule 60, which will be the relevant CA for the TPM: if we identify the empty and occupied sites of the CA with the up and down spins of the TPM, then Rule 60 is the same as the condition that the product of spins in \er{Hsq} is one, thus maximising the local energy and minimising \er{Hsq}. Figure \ref{fig:Ruleplots}(b) also shows the closely related Rule 90.  

\begin{figure}
    \centering
    \begin{subfigure}[b]{0.3\columnwidth}
      \centering
      \includegraphics[width=\linewidth, height=26mm]{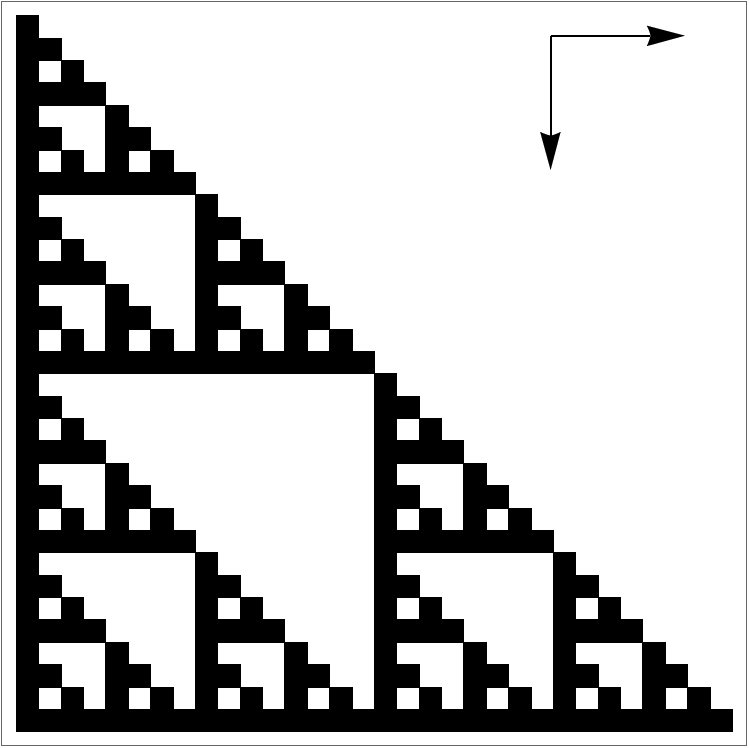}
      \caption{}
      \label{fig:rule60_evolution_singlesite}
    \end{subfigure}%
    \hfill
    \begin{subfigure}[b]{0.3\columnwidth}
      \centering
      \includegraphics[width=\linewidth, height=26mm]{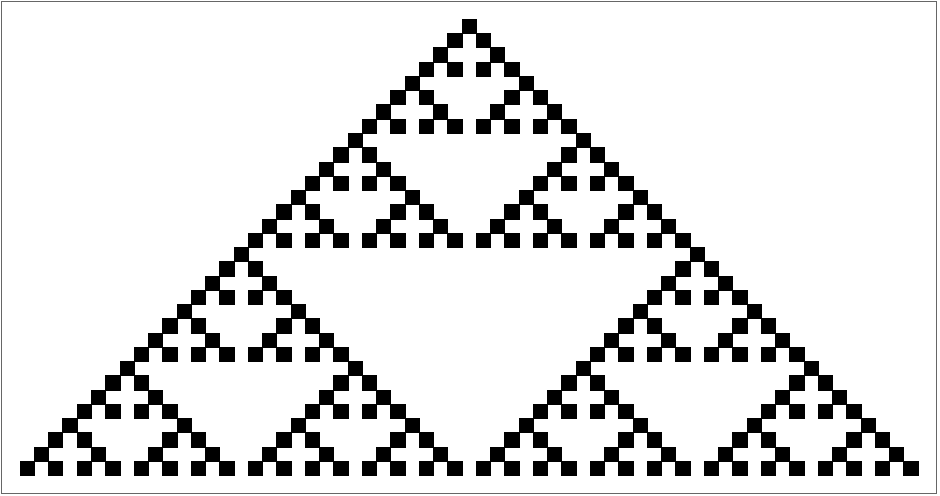}
      \caption{}
      \label{fig:sierpinskitriangle}
    \end{subfigure}%
    \hfill
    \begin{subfigure}[b]{0.3\columnwidth}
      \centering
      \includegraphics[width=\linewidth, height=26mm]{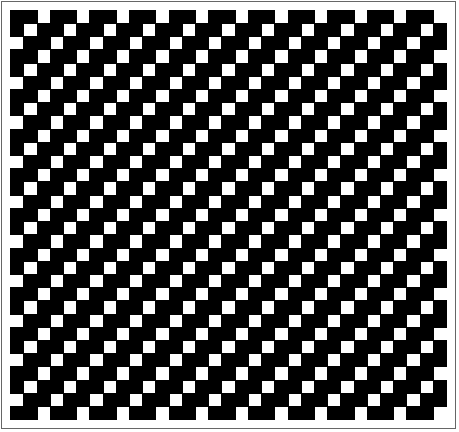}
      \caption{}
      \label{fig:rule60_evolution_periodicstructure}
    \end{subfigure}%
    \caption{(a) Evolution from a single site for Rule 60. (b) Same for Rule 90. (c) A stable cycle generated by Rule 60.}
    \label{fig:Rules}
\end{figure}

Figure \ref{fig:Rules}(a,b) shows the patterns generated for Rules 60 and 90 starting from an initial single seed. Note that these depend on the boundary conditions (for a generic analysis on cellular automata with periodic boundaries, see Ref.~\cite{1988_Jen_cylindrical}). For example, Fig.~\ref{fig:Rules}(b) shows Rule 90 for $L = 64$ and periodic boundaries: the timestep after the last one shown will take the CA to the trivial, empty configuration; in contrast for a length $L = 63$ the Sierpinski fractal \cite{1915_Sierpinski} will continue. Similarly, time evolution will bring in one timestep the pattern of Fig.~\ref{fig:Rules}(a) to the trivial one for PBC and $L = 32$, but the fractal shape would continue for open boundary conditions (OBC). 

Focusing for concreteness on Rules 60 and 90, the CA evolution can be written as a polynomial (or generating function, see Ref.~\cite{1984_Wolfram}). More concretely, any row of the lattice of length $N$ can be expressed as a characteristic polynomial, 
\begin{equation}
    A^{(t)}(x) = \sum_{i = 1}^{N} \alpha_{i - 1}^{(t)} x^{i - 1}.
\end{equation} 
For elementary CA, the value of a site at time t is given by a function $f$,
\begin{equation}
    \alpha_{i}^{(t)} = f\left(\alpha_{i - 1}^{(t - 1)}, \alpha_i^{(t - 1)}, \alpha_{i + 1}^{(t - 1)}\right).
\end{equation}
For additive CA this function calculates the value of $\alpha_{i}^{(t)}$ as a linear combination of the values of the sites in its neighbourhood. For example, for rule 60
\begin{equation}
    f_{60}(x) = 1 + x,
\end{equation}
with the generic evolution then obeying \cite{1984_Wolfram} 
\begin{equation}
    A^{(t)}(x) = f_{\mathrm{CA}} A^{(t - 1)} (x) \, \mod(x^L - 1).
\end{equation}
Note that multiplication by a polynomial, $x^j$, translates the value of a given site j-times \cite{1984_Wolfram}.

Given a configuration at timestep $t$ for the evolution of any given rule, the configuration at timestep $t + 1$ will be its successor and the one at $t - 1$ its predecessor. For Rules 60 and 90, some generic results follow based on Ref.~\cite{1984_Wolfram}:
\begin{itemize}
    \item 
    There are no predecessors for configurations with an odd number of sites being equal to 1 for both rules. For Rule 90,  $2^{L-1}$ configurations for a system of size $L$ cannot be reached if $L$ is odd and $3 \times 2^{L-2}$ if $L$ is even. For Rule 60,  $2^{L-1}$ configurations cannot be reached for any system size. This means that rows with an odd number of ones can only occur as initial states for any given time evolution for Rule 60.
    \item
    Similarly, there may exist configurations that can be reached in one timestep by more than one predecessors. More specifically, given a configuration with at least one predecessor, there are 2 predecessors if the length $L$ is odd and 4 if it is even for Rule 90, while there are always 2 for Rule 60.
\end{itemize}
Regarding the length of cycles for Rule 90, more results can be found in Ref.~\cite{1984_Wolfram}. For Rule 60, which is most relevant for the TPM, we discuss the cycle lengths and their multiplicities below.

\begin{figure*}
    \includegraphics[width=\textwidth]{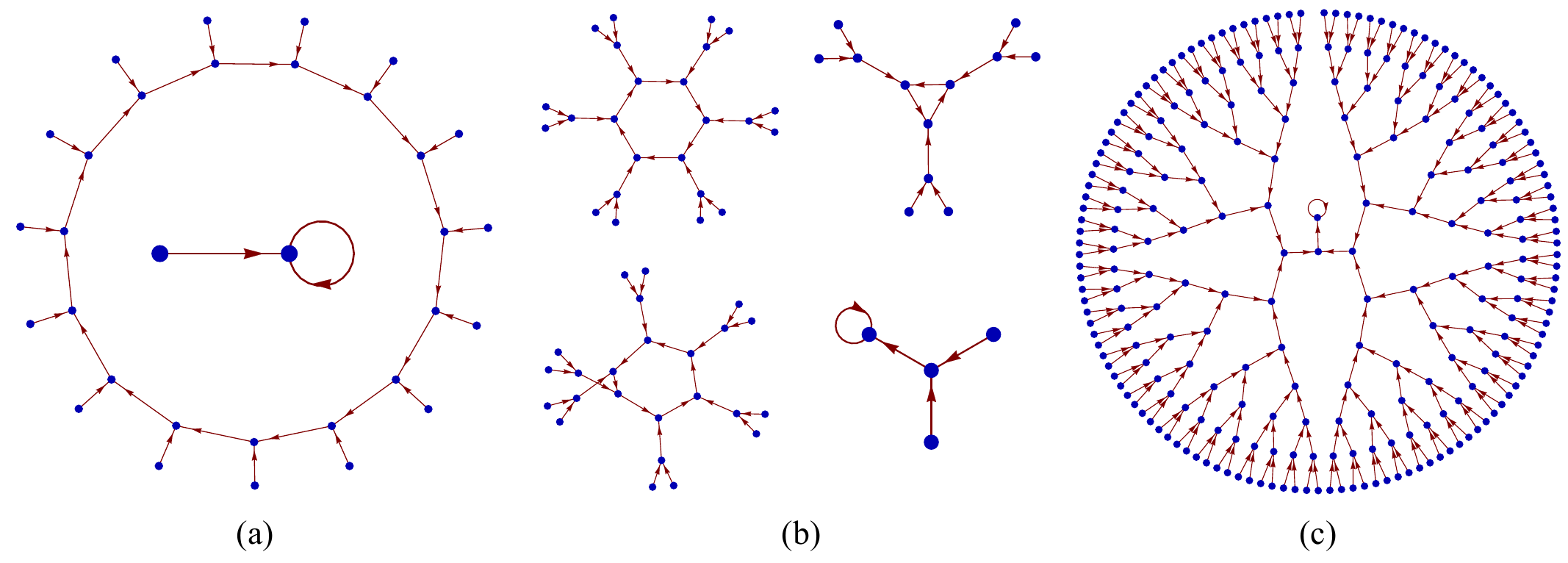}
     \caption{The fixed point and limit cycles of Rule 60. Filled circles indicate states of the CA and the arrows the flow under the dynamics. 
     (a) For $L=5$ there is a cycle of length $\mathcal{C}=15$ and the fixed point (which flows into itself). 
     (b) For $L=6$ there are two cycles of period $\mathcal{C}=6$, a cycle of period $\mathcal{C}=3$ and the fixed point. 
     (c) For $L=8$, there is the fixed point and no cycles.
     }
     \label{fig:Cycles}
\end{figure*}

\begin{table}
    \centering
    \includegraphics[width=.37\textwidth, height=21cm]{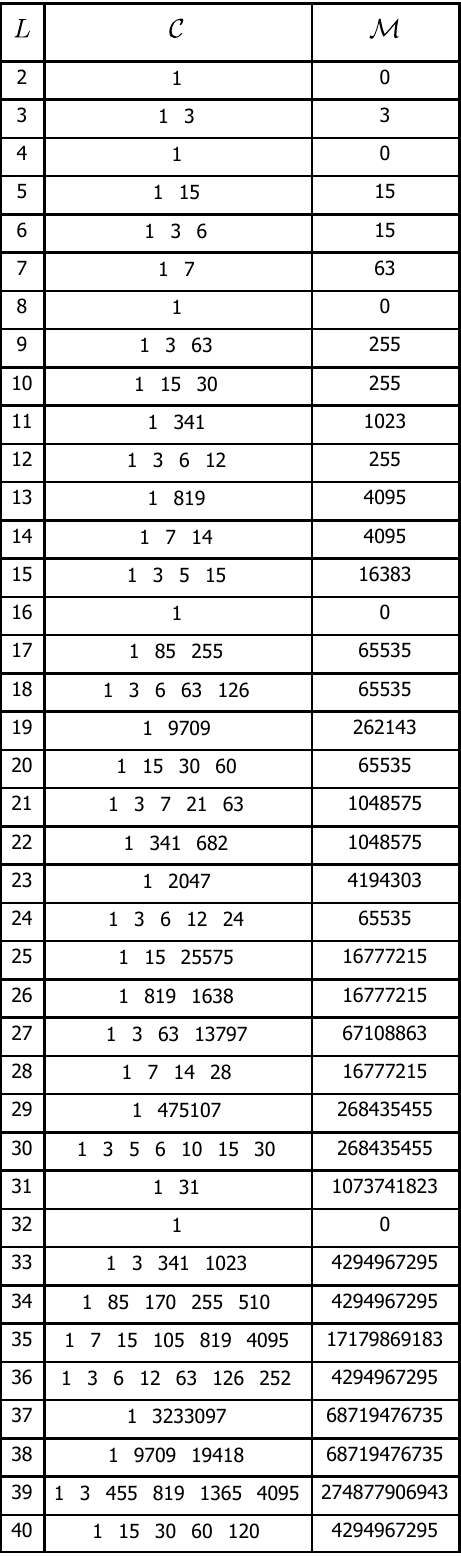}
    \caption{\label{fig:Cycle_Lengths_up_to_40} Cycle lengths for Rule 60. The second column indicates the cycle lengths and the third the
     total number of periods for Rule 60 (each period of length $\mathcal{C}$ is counted $\mathcal{C}$-times), assuming that the least common multiple 
     of the periods for each given $L$ divides $M$, $\lcm \mathcal{C} | M$.
     The last column is constructed based on the number of predecessors for all configurations for Rule 60, given a length $L$.
    }
\end{table}

\subsection{Periodic orbits of Rule 60}

The iterated map
\begin{equation}
    D_n(\mathbf{x}) = (\abs{x_1 - x_2}, \abs{x_2 - x_3}, ..., \abs{x_L - x_1})
\end{equation}
for $D_n : \mathbb{Z}^n \rightarrow \mathbb{Z}^n$ and $\mathbf{x} = (x_1, x_2, ..., x_L)$ is known as the Ducci map (also known as rule 102 in Wolfram's notation \cite{1983_Wolfram,1984_Wolfram}, the mirror image of Rule 60, e.g.\ Ref.~\cite{Mathworld_weisstein_rule_nodate}). Ducci's map (and therefore Rule 60) exhibits periodic orbits or evolves to a fixed point depending on $L$ being a power of two or not. For $L=2^k$, any initial state evolves to the trivial state of zeros \cite{2000_andriychenko_iterated}, while for $L \neq 2^k $ it evolves to a richer attractor structure \cite{1990_Ehrlich}.

In $\mathbb{F}_2$ and by using periodic boundary conditions, the Ducci map can be brought into matrix form as follows,
\begin{equation}
    D_n \, \mathbf{x} = \left( (x_1 + x_2), (x_2 + x_3), \cdots, (x_L + x_1)\right) \mod{2}, 
\end{equation}
where
\begin{equation}
    D_n = \begin{bmatrix}
    1 & 1 & 0 & \ldots & 0 & 0 & 0 \\
    0 & 1 & 1 & \ldots & 0 & 0 & 0\\
    \vdots & \vdots & \vdots & \ddots & \vdots & \vdots &\vdots  \\
    0 & 0 & 0 & \ldots & 1 & 1 & 0 \\
    0 & 0 & 0 & \ldots & 0 & 1 & 1 \\
    1 & 0 & 0 & \ldots & 0 & 0 & 1 \\
\end{bmatrix}.
\end{equation}
The  matrix $D$ can be expressed as
\begin{equation}
    D = I + S_{L}
\end{equation}
with $S_L$ the left shift map \cite{1990_Ehrlich,2005_Calkin}. 
It is easy to check that 
\begin{equation}
    D^{2^k} = 
    \left(
        I + S_L 
    \right)^{2^k} 
    = I + S_{L}^{2^k}
    = I + I = 0 \mod{2},
\end{equation}
with $k \in \mathbb{Z}$ and, thus, a system that has $L$ a power of 2 ends up in the trivial configuration. For Rule 60 the corresponding matrix is ${D}^\intercal$.

In order to obtain the cycles of Rule 60, we need the following definitions and results for Rule 102 from Refs.~\cite{1993_Stevens,1999_Stevens,1990_Ehrlich,1988_Jen_cylindrical,2003_Chamberland_Thomas,2005_Calkin,2006_Misiurewitz,2006_Thomas_stevens_Lettieri}:
\begin{itemize}
    \item
    For any given CA, each array of sites can be thought of as a vector, $\mathbf{v}$.
    For any such vector $\mathbf{v}$ in $\mathbb{F}_2$, its order is defined through the monic polynomial, which satisfies
    $\mu_{\mathbf{v}}(D) \mathbf{v} = 0$. 
    \item
    The order of the minimal annihilating polynomial, $\ord{\mu_{\mathbf{v}}(\lambda)}$, is equal to the smallest natural number $n$, such that 
    $\mu_{\mathbf{v}}(\lambda)|\lambda^n - 1$. 
    If $\mu_{\mathbf{v}}(0) = 0$, then $\mu_{\mathbf{v}}(\lambda) = \lambda^k \tilde{\mu}_{\mathbf{v}}(\lambda)$ with $\tilde{\mu}_{\mathbf{v}}(0) \neq 0$ and $k \in \mathbb{N}$.
    Thus, the order of $\mu_{\mathbf{v}}$ is $\ord{(\mu_{\mathbf{v}})} = \ord{(\tilde{\mu}_{\mathbf{v}})}$.

    \item
    Assuming that $\mu_{\mathbf{v}}(\lambda) = \lambda^k \tilde{\mu}_{\mathbf{v}}(\lambda)$ with $k\geq 0$, then the $k$-th successor of 
    $\mathbf{v}$ belongs to a cycle of length $c = \ord{\mu_{\mathbf{v}}}$.
    This applies to a vector $\mathbf{v}$ of any positive integer $L$ and any linear map (or CA rule).
\end{itemize}

The minimal annihilating polynomial for the Ducci map was calculated in Refs.~\cite{2005_Calkin,2006_Misiurewitz}, based on the characteristic polynomial of the matrix $D$. It was found that
\begin{equation}
    \mu_n(\lambda) = p_n(\lambda) = (1 + \lambda)^L +1.
\end{equation}

We are now in a position to obtain the attractor structure of Rule 60 by following Ref.~\cite{1999_Stevens} (specifically ``Principle C''). We decompose the minimal annihilating polynomial into the product of its irreducible polynomials, $\pi_i(\lambda)$. We call their polynomial powers 
$b_i$. Thus, 
\begin{equation}
    \mu_{\mathbf{v}}(\lambda) = \prod_{i=1}^{m} \pi_i(\lambda)^{b_i},
\end{equation}
and 
\begin{equation}
    \ord{\mu_{\mathbf{v}}(\lambda)} = r 2^t,
\end{equation}
where $r$ is the least common multiple of $\ord{\pi_i}$ and $t$ the smallest integer satisfying $2^t \geq \max{(b_1, b_2, \dots, b_m)}$.

Figure \ref{fig:Cycles} illustrates the different scenarios for cycles as a function of $L$. We show three different sizes, $L=5,6,8$, small enough to be able to visualise the network of states. Configurations of the CA are identified by blue circles, and arrows indicate to which configurations they evolve to under the CA dynamics. Figure \ref{fig:Cycles}(a) shows $L=5$: here there is one fixed point to which one other state evolves to (shown at the centre of the figure), and one limit cycle of period $\mathcal{C}=15$, to which all other configurations flow. Figure \ref{fig:Cycles}(b) shows a more general situation of multiple distinct cycles for the case of $L=6$: there are two cycles of period $\mathcal{C}=6$, one cycle of period $\mathcal{C}=3$ and one fixed point. For the case of $L$ a power of 2, as shown in Fig.~\ref{fig:Cycles}(c) for $L=8$, there is a unique fixed point and all states evolve towards it.

\subsection{Classical ground states of the TPM}

From the analysis above for Rule 60, we can enumerate all the minimum energy configurations of the TPM, cf.\ \er{Hsq}. The classical ground states for a system of size $N = L \times M$ are: (i) the state with all spins up, corresponding to the fixed point of Rule 60, for any value of $L$ and $M$; (ii) two-dimensional spin configurations that correspond to periodic trajectories of Rule 60, that is, CA trajectories starting from any of the states of a limit cycle, for all limit cycles whose period is contained an integer number of times in $M$---this occurs only for certain combinations of $L$ and $M$ (never if $L$ or $M$ is a power of two). 

We show the relevant Rule 60 information in Table \ref{fig:Cycle_Lengths_up_to_40} for up to $L=40$. Under the column ${\cal C}$ we give the distinct periods of the limit cycles. In the column labelled by ${\cal M}$ we give the corresponding degeneracy of classical ground states of the TPM, apart from the uniform spin-up state, given that $\lcm \mathcal{C} | \cal M$. For example, for $L=15$, there is one cycle of length 3, three cycles of length 5, and 1091 cycles of length 15, which means $1+{\cal M} = 1 + 3 + 3 \times 5 + 1091 \times 15 = 16384$ distinct two-dimensional spin configurations that minimise the energy in a TPM of size $N=15 \times M$ as long as $15 | M$. In contrast, for $L=15$, if $5|M$ but $3 \nmid M$ and $15 \nmid M$, then there are $1+15$ different ground states. 
\section{Phase transition in the quantum TPM}
\label{PT}

As we will now show, the set of minimum energy configurations and the associated symmetries of the classical TPM, as obtained from the Rule 60 CA, determine the properties of the ground state phase transition in the quantum TPM, \er{QTPM}.

\subsection{Symmetries of the QTPM}

Like its classical counterpart, the QTPM with PBC has full translational invariance. The symmetries of the QTPM will then be deduced by the results of Sec.~\ref{CA_section}. 

\begin{figure}
    \centering
    \includegraphics[width=\columnwidth]{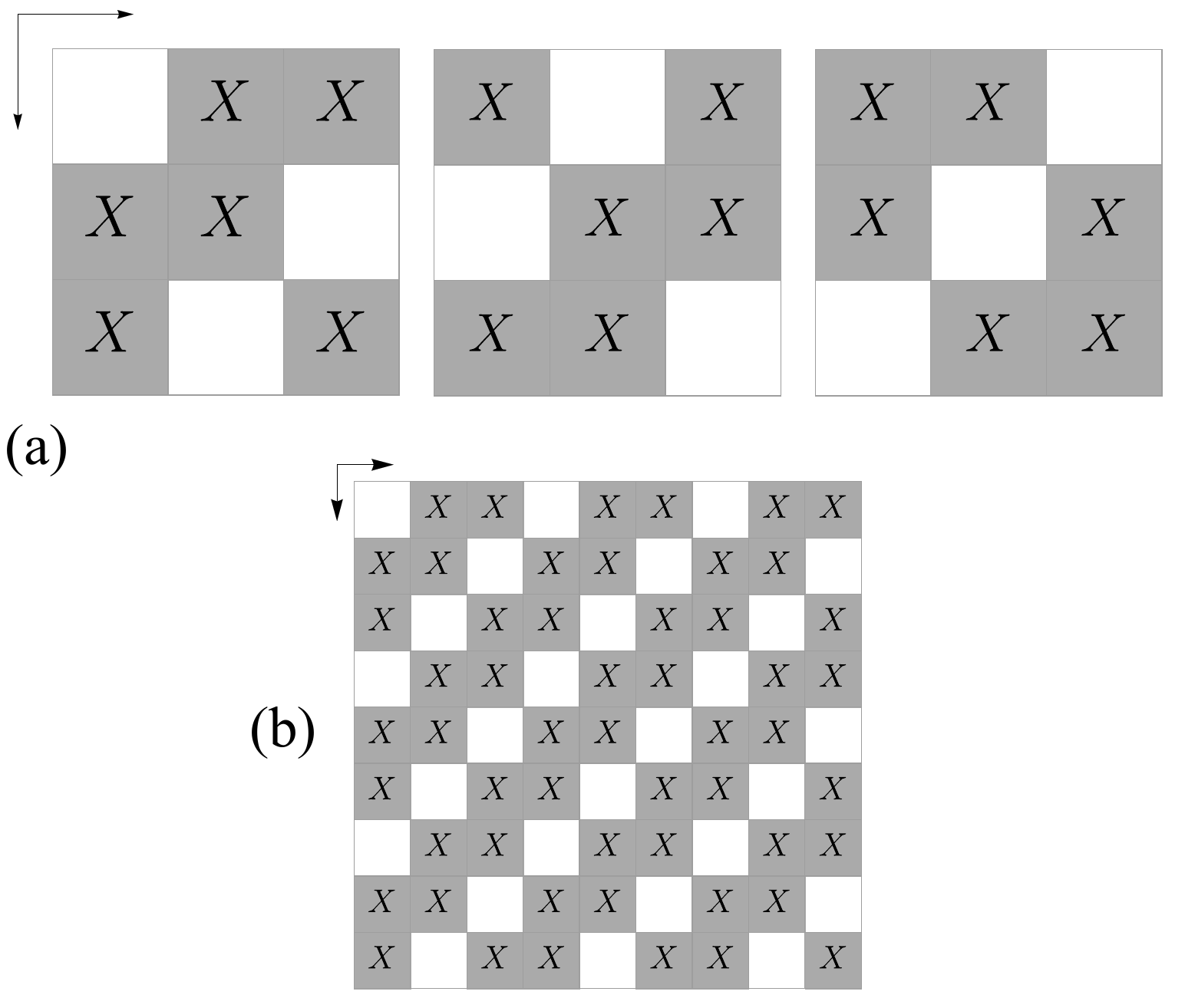}
     \caption{(a) Symmetry operators for the $3\times 3$ QTPM with periodic boundaries. 
     (b) One of the symmetry operators of the $9\times 9$ QTPM.}
     \label{fig:QTPM_3x3_symmetries}
\end{figure}

For systems with dimensions $N=L \times M$, which can accommodate non-trivial cycles of Rule 60, the symmetries of the corresponding QTPM easily follow. Consider as a simple example the case of $N = 3\times 3$ with periodic boundaries in both dimensions. From Table~I, we see that $L=3$ has one cycle of period 3. This means that for $M=3$ there are three non-trivial symmetries, given by the operators 
\begin{align}
    \label{G1}
    G_1 &= X_{1,2} X_{1,3} X_{2,1} X_{2,2} X_{3,1} X_{3,3} \\
    G_2 &= X_{1,1} X_{1,3} X_{2,2} X_{2,3} X_{3,1} X_{3,2} \\
    \label{G3}
    G_3 &= X_{1,1} X_{1,2} X_{2,1} X_{2,3} X_{3,2}  X_{3,3}
\end{align}
see Fig.~\ref{fig:QTPM_3x3_symmetries}(a). Note that translational invariance plays a crucial role in the determination of the symmetries. In the example above $G_2=\mathcal{T}_x G_1 \mathcal{T}_x^{-1}$ and $G_3=\mathcal{T}_x G_2 \mathcal{T}_x^{-1}$, where $\mathcal{T}_x$ is the translation operator in the $x$ direction. These symmetries alongside the identity form the Klein group $\mathcal{K}_4$, which, in turn, is isomorphic to $\mathbb{Z}_2 \otimes \mathbb{Z}_2$. 

This approach generalises to other system sizes, and the symmetries are products of $\mathcal{K}_4$. For example, the symmetries of the $N=5\times 15$ and the $N=6 \times 6$ systems form the group $\mathcal{K}_4 \otimes \mathcal{K}_4$. Since the symmetry operators are products of the $X$ Pauli matrices, cf.~\ers{G1}{G3}, they commute with the transverse field term in \er{QTPM}, and, therefore, are symmetries of the QTPM for all values of $J$ and $h$. They are subsystem symmetries of the global $\mathbb{Z}_2$ symmetry, similar in nature to type-II fracton models \cite{2019_Nandkishore_Fractons,2020_Pretko}.

\subsection{The character of the quantum phase transition}

The character of the phase transition for the QTPM is now easy to predict based on the information of its symmetries for a given system size. By this we mean: given a sequence of increasing system sizes with periodic boundaries, all having the same or monotonically increasing number of symmetries, the progressively sharper finite size crossovers will be indicative of an eventual phase transition in the large size limit whose character---first-order or continuous---will be guided by the underlying symmetries of the given system sizes in the sequence. In a system of size $N = L \times M$  as these symmetries depend on the precise values of both $L$ and $M$, this analysis has to be done carefully. In the next section we show numerical evidence for the general considerations we give here. 

First consider the system sizes $N = L \times M$ such that the underlying Rule 60 has only one fixed point, as described previously. Then, the QTPM has no non-trivial symmetries. In the limit $J \gg h$, there is a single ground state corresponding to the all-up state, and a vanishing number of excited plaquettes, as the classical TPM has a single minimum. In the opposite limit, $J \ll h$, the ground state is paramagnetic, with spins aligned in the $x$ direction, with an explicit $\mathbb{Z}_2$ symmetry for its ground state, corresponding to a high density of excited plaquettes. In the limit of large $N$, we expect the quantum phase transition at the self-dual point $J=h$ to be first-order, due to the explicit breaking of the $\mathbb{Z}_2$ symmetry. We verify our conjecture with numerical simulations in the next section.

A second scenario is that of system sizes where the underlying Rule 60 cycles give rise to non-trivial symmetries in the QTPM. In this case, in the limit $J \ll h$, the ground state is the same paramagnetic one as before, invariant under a global $\mathbb{Z}_2$ symmetry. However, for $J \gg h$ there exists spontaneous breaking of the symmetries emerging from Rule 60. This case, therefore, has characteristics of a first-order phase transition, as before, with additional spontaneous symmetry breaking (SSB). 
    This case is reminiscent of the SSB in the transverse field Ising model (TFIM). There, the two phases are the quantum paramagnet and the classical ferromagnet which is doubly degenerate. In the thermodynamic limit, there is a continuous quantum phase transition at the quantum critical point since in this situation both terms respect the global $\mathbb{Z}_2$ symmetry. The SSB originates in the freedom in choosing between the two degenerate classical ground states \cite{sachdev_QPTbook}. However, it is important to emphasize that in the case of the QTPM the SSB scenario accompanies a first-order quantum phase transition.

This is similar to the first-order phase transitions observed in kinetically constrained models \cite{2021_Luke} and suggests that, for local observables, the phase transition will appear first-order; for example, in a discontinuous jump in the excited plaquette density, 
\begin{equation}{\label{Mzzz}}
    M_{zzz} = \frac{1}{N}\sum_{\{i, j, k\} \in \triangledown} Z_i Z_j Z_k.
\end{equation}
Appropriate operators will quantify the symmetry breaking of the degenerate classical ground states.

The choice of these operators depends on the size and specific lattice dimensions of the system. Consider, for example, the case of $L = 3$ and $M = 3k$ 
with $k\in \mathbb{N}$, where we know from Rule 60 that there is a single fixed point and the three non-trivial ground states. For the trivial ground state this operator is just the magnetisation $M_z = \frac{1}{N} \sum_{i}^{N} Z_i $. To detect the symmetry breaking for the other three states, we can define the three operators $\tilde{M}_z^{m} = \frac{1}{N} \sum_{i}^{N} (-1)^{n_i} Z_i $, where $n_i=1$ if the spin $i$ is flipped for a state of the cycle $m$ of Rule 60, and $n_i=0$ otherwise. For example, for the state associated to \er{G1}, we have 
\begin{align}{\label{OPLuke}}
    \tilde{M}_z = 
    \frac{1}{N} & \left(  Z_{1,1} - Z_{1,2} - Z_{1,3} - Z_{2,1} - Z_{2,2} +Z_{2,3}  
    \right. 
    \nonumber \\
    &
    \left. 
    - Z_{3,1}  + Z_{3, 2} - Z_{3,3} \right).
\end{align}
Note that when there are multiple non-trivial ground states connected by translations, there will be no single operator taking the form of a sum of local terms for which it will be possible to discern between them. For example, in the $N = 3 \time 3k$ case, $M_z$ will only distinguish between the trivial ground state and the 3-fold degenerate ones, while the operators $\tilde{M}_z^{m}$ will be able to distinguish only one of the non-trivial ground states.

\subsection{Numerics}{\label{numerics}}

We now provide evidence for the general observations above from numerical simulations. For small systems we use Exact Diagonalization (ED) 
\cite{2010_Sandvik,book_computational_many_body,book_strongly_correlated_methods}, allowing the study of system sizes up to 28 sites. For larger systems we estimate the properties of the ground states using two different approaches. The first of these is Matrix Product States (MPS) \cite{2011_Schollwock,2021_Cirac_MPSreview,2020_Itensor,2014_Orus_Review}, which we ``snake'' around the 2D lattice, and optimize with the 2D Density-Matrix Renormalization Group (DMRG) \cite{1992_White, 2011_Stoudenmire}. By employing a bond dimension up to $D = 1000$, we are able to reliably estimate the ground state properties for system sizes on the square lattice for up to $N=16\times 16$. Time and memory constraints hinder progressively the convergence in the paramagnetic phase, where $J \sim h$. As discussed below, we are also able to apply MPS to cylindrical systems, which can be considered to be quasi-1D, allowing us to reach much larger sizes than in the case of square geometries.

To confirm the results of 2D DMRG, we also employ Quantum Monte Carlo (QMC) methods. In particular, we use the {\em continuous-time expansion}  (ctQMC) \cite{1996_Beard}, with local spin updates which re-draw the entire trajectory of a single spin, subject to a time-dependent environment, where the trajectories of unmodified spins are considered to act as a ``heat bath'', e.g., see Refs.~\cite{2008_Krzakala, 2012_Mora}. We run our simulations with an inverse temperature of $\beta = 128$, which we find to be large enough to converge to the ground state \cite{2023_Causer}.

\subsubsection{First-order transition}

As explained above, when the underlying Rule 60 has a single fixed point, and the classical TPM a single energy minimum with all spins up, we thus expect the phase transition of the QTPM to be first-order. 

\begin{figure}
    \centering
    \includegraphics[width=\singlefigure]{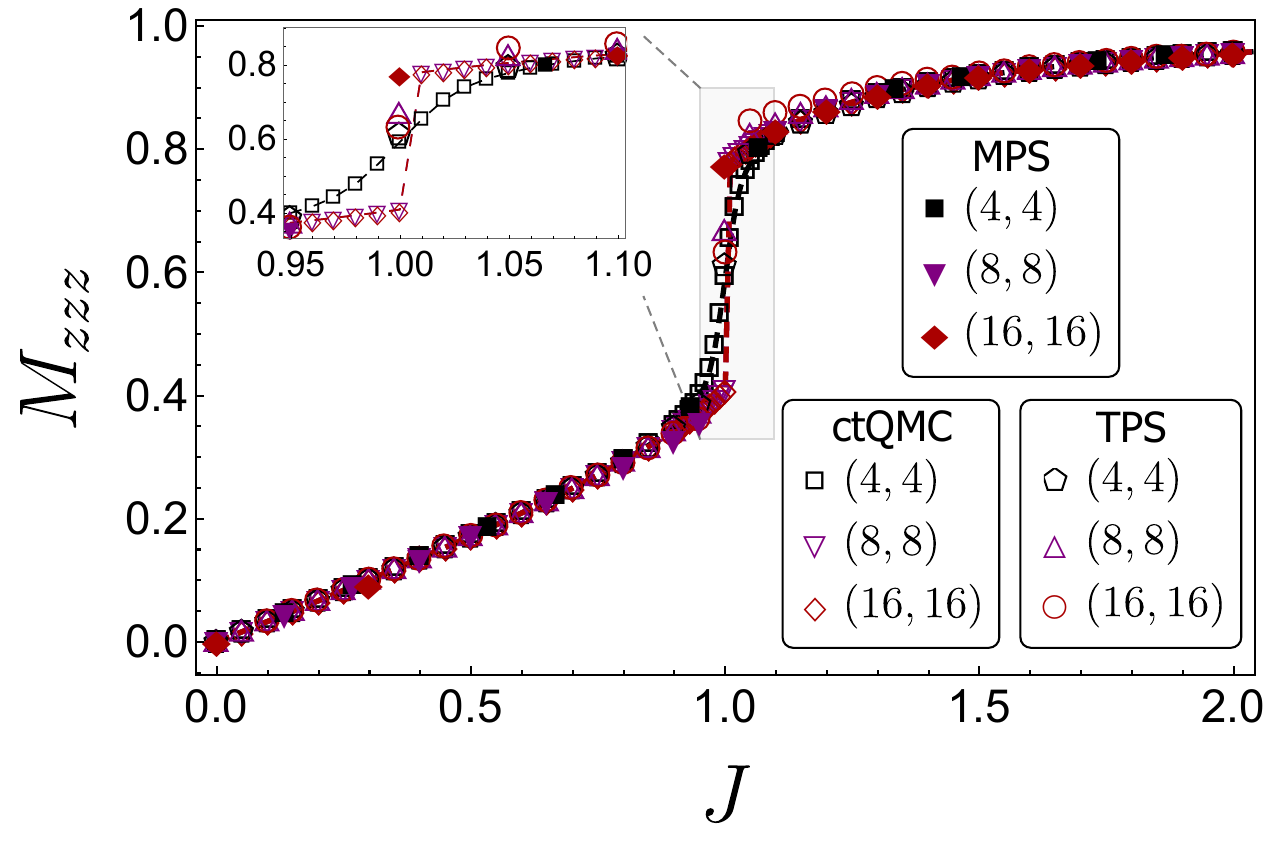}
    \caption{\label{fig:comparison_with_Loredana} 
    Normalised three-spin correlator, \er{Mzzz},  in the QTPM as a function of $J$ for fixed $h=1$, with $N = L \times L$ with $L$ a power of two. 
    We compare results from MPS and ctQMC obtained here with results from Ref.~\cite{2020_Vasiloiu}. 
    The numerical data indicates a first-order transition at $J=h$.}
\end{figure}

Figure \ref{fig:comparison_with_Loredana} shows results for $N = L \times L$ with $L = 4, 8, 16$. We show that our MPS and ctQMC results coincide with the large deviations results of Ref.~\cite{2020_Vasiloiu}, obtained via transition path sampling (TPS). Further, Fig.~\ref{fig:comparison_with_Loredana} depicts the plaquette density $M_{zzz}$ as a function of the coupling $J$ for fixed transverse field $h=1$. The data indicates a first-order transition at $J=h$, as expected. For $J>1.0$, we see small deviations in the TPS results, due to the extra field used for the acquisition of this data in Ref.~\cite{2020_Vasiloiu}. Similar issues are observed for ctQMC close to the $J = 1.0$ point, where the single spin updates do not allow for the collective effects necessary to move between phases.

\begin{figure*}
    \centering
    \begin{subfigure}[b]{0.3\textwidth}
        \includegraphics[width=54mm, height=39mm]{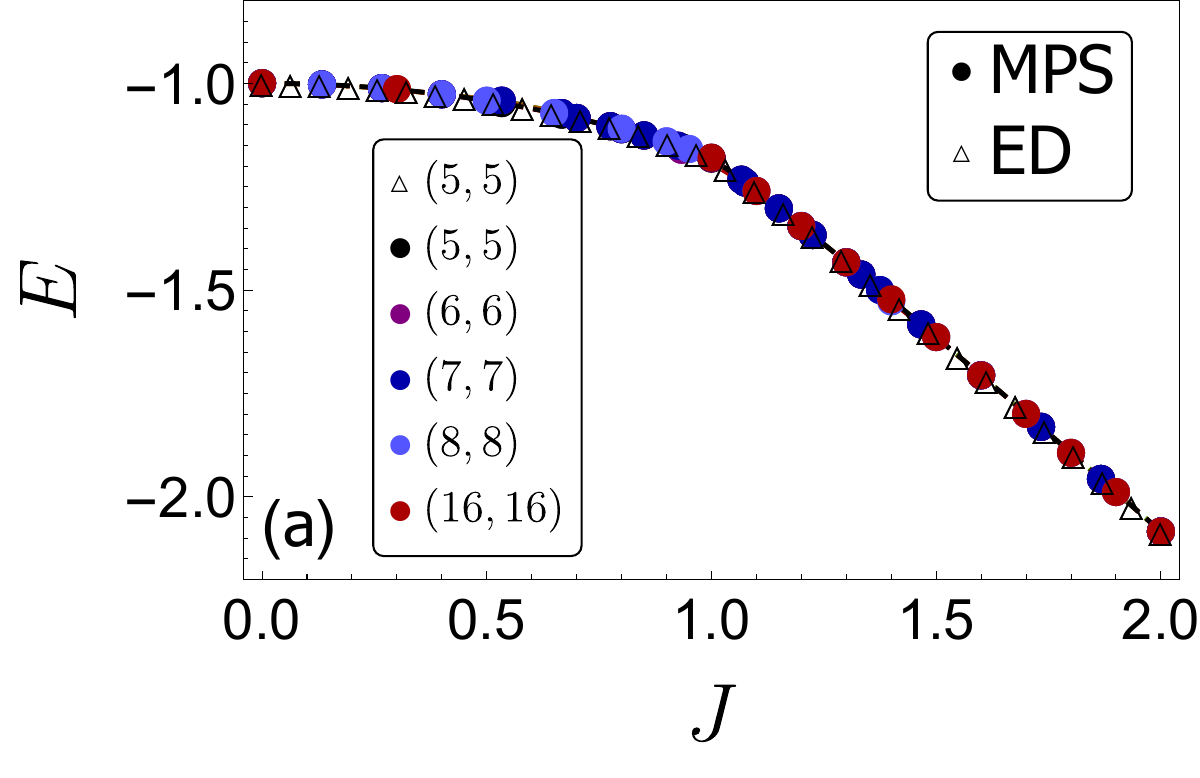}
     \end{subfigure}
     \begin{subfigure}[b]{0.3\textwidth}
        \includegraphics[width=54mm, height=39mm]{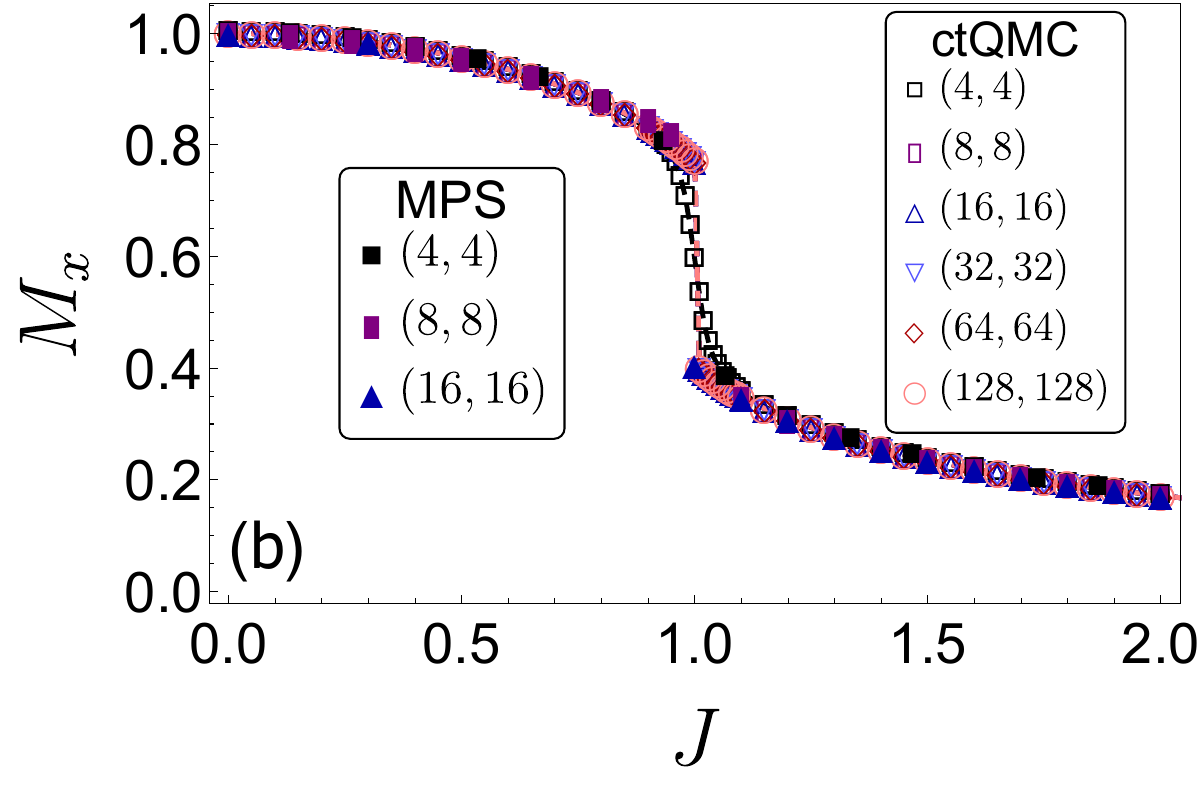}
     \end{subfigure}
     \begin{subfigure}[b]{0.3\textwidth}
        \includegraphics[width=54mm, height=40mm]{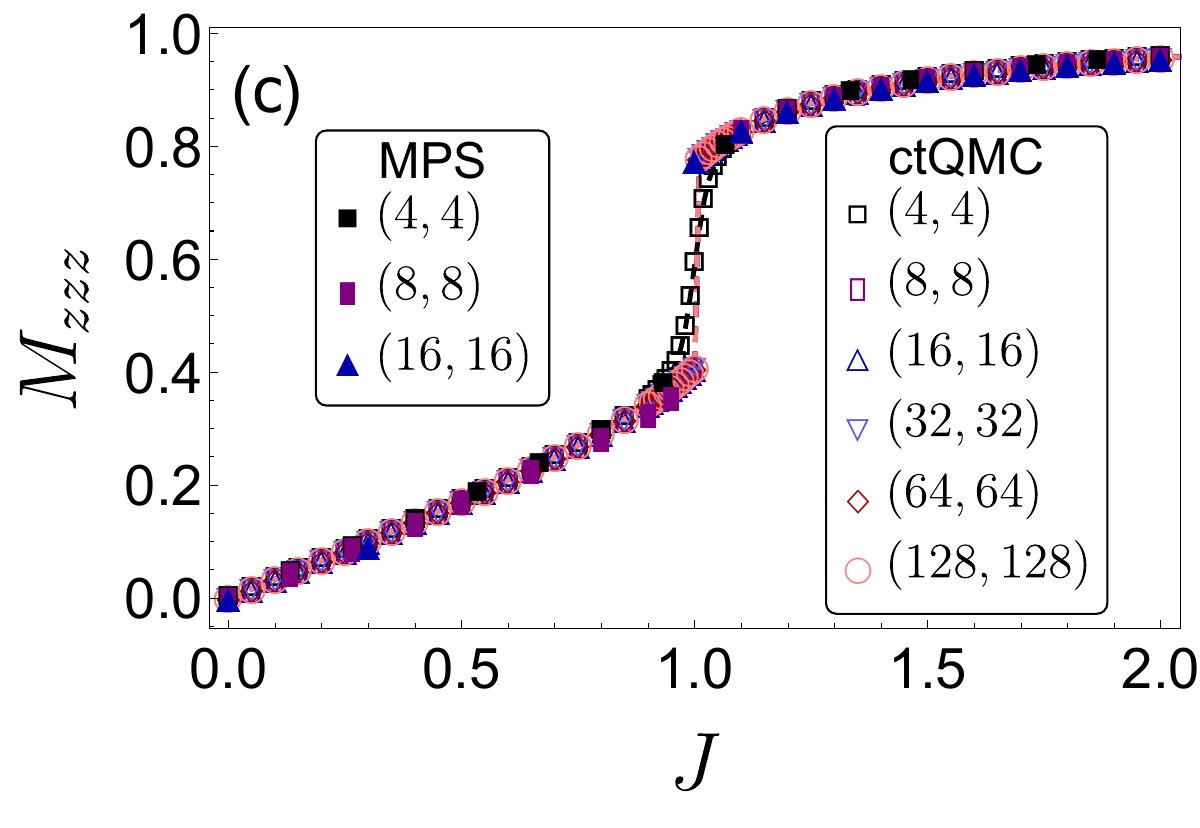}
     \end{subfigure}
     \caption[PBC, square]{First-order transition in the QTPM for systems of size $N = L \times L$.
        (a) The normalised by the system size ground state energy as a function of $J$ at fixed $h=1$. Open symbols are results from ED, filled symbols from numerical MPS. 
        (b) Transverse magnetisation as a function of $J$. In this case the open symbols are from ctQMC. 
        (c) Average three-spin interaction as a function of $J$.}
     \label{fig:MPS_ED_ctQMC_square}
\end{figure*}

\begin{figure*}
    \centering
    \begin{subfigure}[b]{0.3\textwidth}
        \includegraphics[width=54mm, height=39mm]{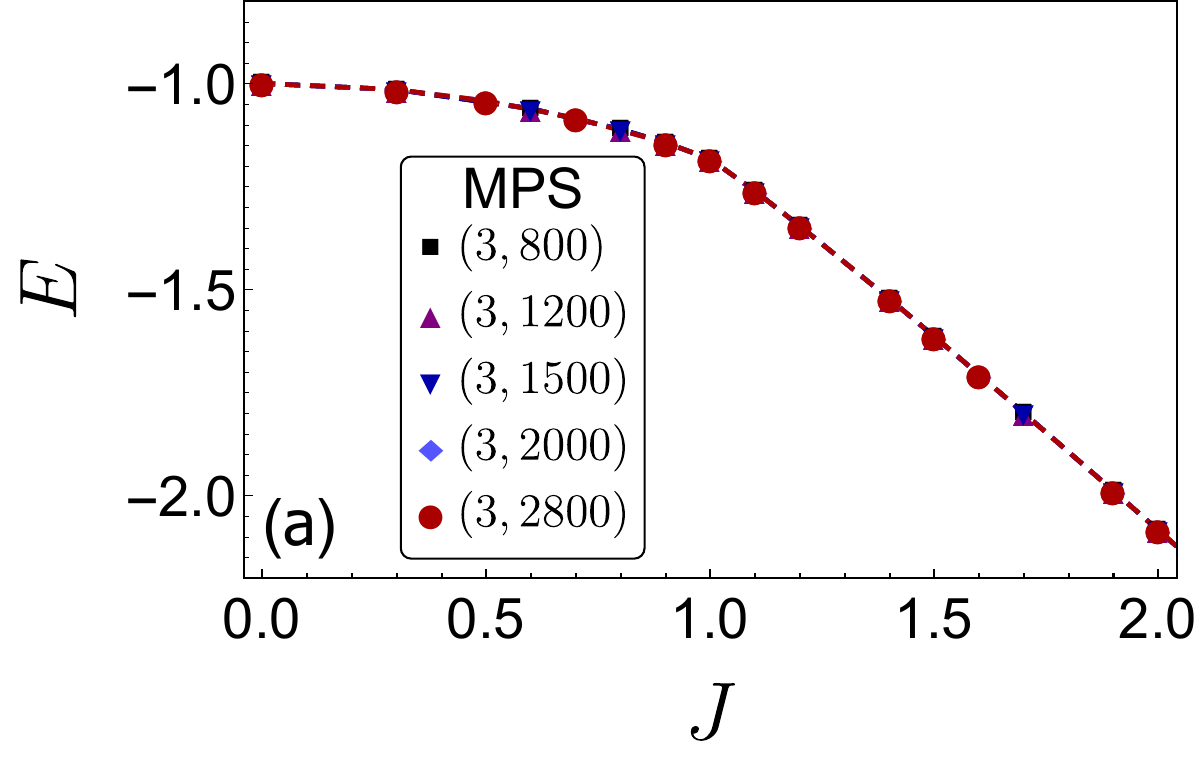}
     \end{subfigure}
     \begin{subfigure}[b]{0.3\textwidth}
        \includegraphics[width=54mm, height=39mm]{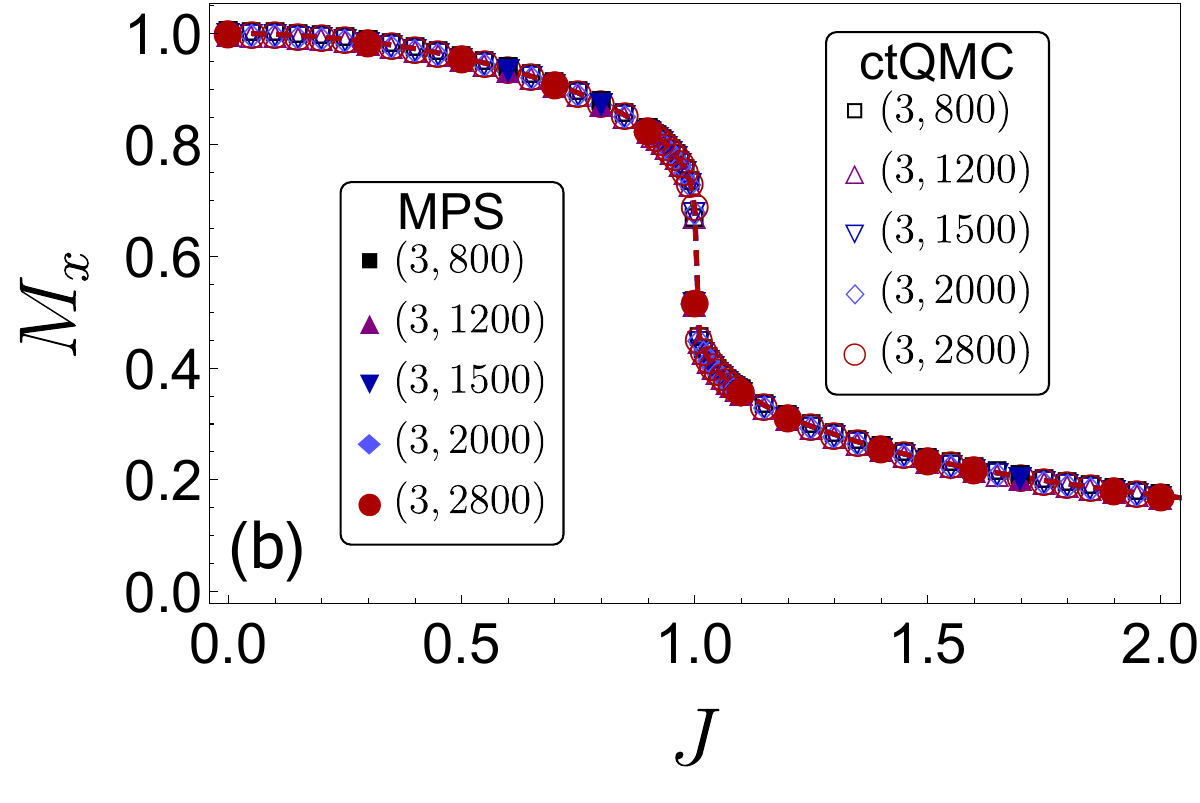}
     \end{subfigure}
     \begin{subfigure}[b]{0.3\textwidth}
        \includegraphics[width=54mm, height=40mm]{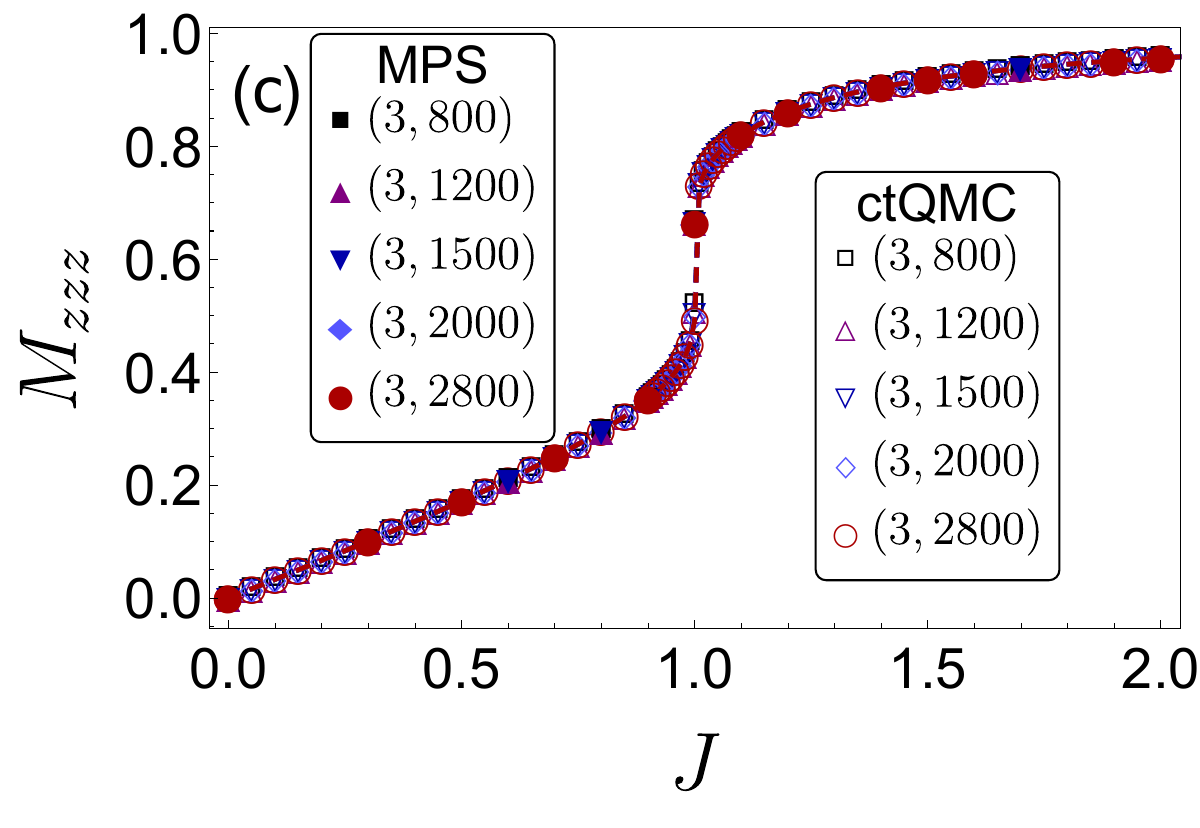}
     \end{subfigure}
     \caption[PBC, rectangular]{Same as Fig.~\ref{fig:MPS_ED_ctQMC_square} but for systems of size $N = 3 \times M$.}
     \label{fig:MPS_ED_ctQMC_rectangular}
\end{figure*}

In Fig.~\ref{fig:MPS_ED_ctQMC_square}, we show the results for several system sizes in a square geometry, $N = L \times L$. Figure ~\ref{fig:MPS_ED_ctQMC_square}(a) shows the ground state energy as a function of $J$ (at $h=1$) for $L=5$ to $16$, obtained from MPS numerics. For the smallest size we also show ED results, which coincide with the MPS ones. The kink near $J=1$ indicates a quantum phase transition. Note that this behaviour is similar in systems with a single classical ground state ($L=5,8,16$) or multiple ones ($L=6,7$), cf.\ Table~\ref{fig:Cycle_Lengths_up_to_40}. In Figs.~\ref{fig:MPS_ED_ctQMC_square}(b,c) we show the average transverse magnetisation, $M_x = \frac{1}{N} \sum_i X_i$, and $M_{zzz}$, respectively, for systems with $L$ a power of two. We get exactly the same results for different system sizes too. Both MPS and ctQMC show clear indications of a first-order transition at $J=1.0$ in both observables.

Figure~\ref{fig:MPS_ED_ctQMC_rectangular} shows similar results in a rectangular geometry, $N = 3 \times M$. For such thin strip systems we can perform MPS more efficiently for larger system sizes than for square geometries. Once again, MPS and ctQMC results coincide, and indicate a first-order transition at $J=1.0$ (although weaker than in the square lattice case, in the sense that the shown discontinuity in the local operators is smaller). Note that these results include not only values of $M$ which are multiples of three, for which there are multiple classical ground states, but also values of $M$ for which a single ground state is found. What we see in this case is that the observables $M_x$ and $M_{zzz}$ are unable to detect changes related to any given classical ground state.

\subsubsection{Symmetry breaking}

\begin{figure*}
    \centering
    \begin{subfigure}[b]{0.45\textwidth}
        \includegraphics[width=8cm, height=5cm]{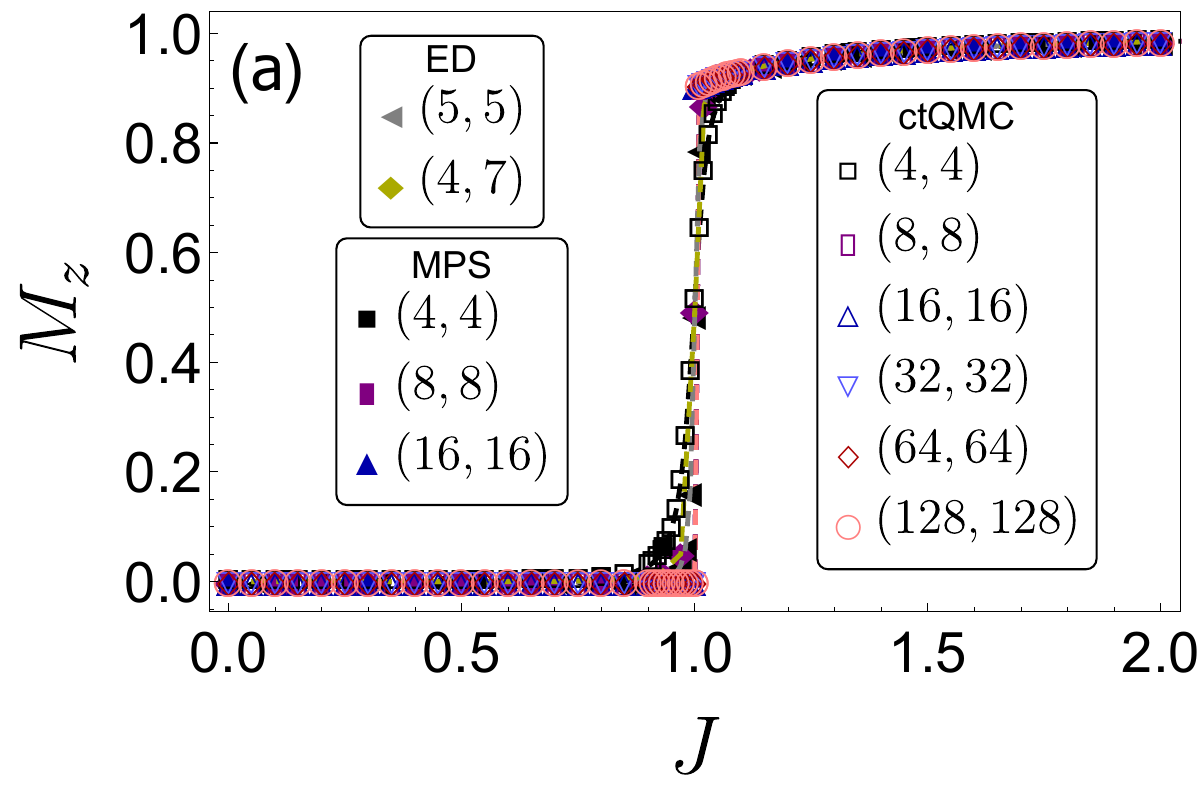}
    \end{subfigure}
    \begin{subfigure}[b]{0.45\textwidth}
        \includegraphics[width=8cm, height=5cm]{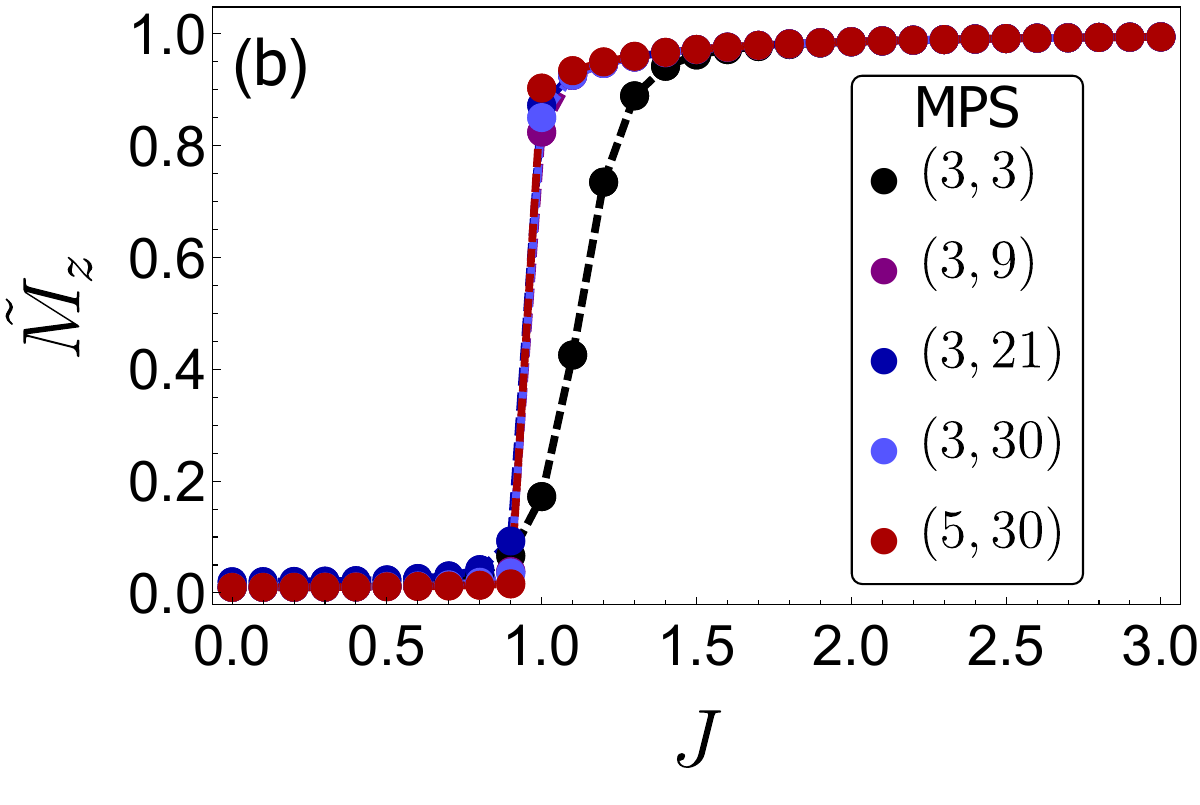}       
    \end{subfigure}
    \caption{\label{fig:OP} (a) Longitudinal magnetisation for systems with no symmetries. (b) Staggered magnetisation for detecting symmetry breaking in systems with multiple symmetries.}
\end{figure*}

In Figs.~\ref{fig:MPS_ED_ctQMC_square} and \ref{fig:MPS_ED_ctQMC_rectangular}, we show the two terms that compete in the Hamiltonian, $M_x$ and $M_{zzz}$. For system sizes where there is one classical ground state and no non-trivial symmetries, the total longitudinal magnetisation $M_z$ can also serve as an order parameter, as it picks up the orientation of the ground state. Figure~\ref{fig:OP}(a) shows that the transition is also clear for this observable for square lattices.

For system sizes in the thermodynamic limit where degeneracies are expected for $J \geq h$, however, $M_z$ is unable to detect the symmetry breaking related to the extra symmetries. For these cases, we need the staggered magnetisations, $\tilde{M}_z^{m}$, such as that for  $N=3 \times 3$ in \er{OPLuke}. Figure~\ref{fig:OP}(b) shows that such operators are able to detect the SSB for these lattices. Note that Fig.~\ref{fig:OP}(b) was obtained through the use of a small symmetry breaking field. This is a standard method for the detection of the symmetry breaking in the ground state of a degenerate quantum model \cite{2021_Lauchli}. As a result, the calculations were performed through the use of a modified Hamiltonian $H = \opcatq{H} - p \tilde{M}_z$, where $p$ is chosen to be small. The detection of the SSB can be similarly preformed for any of the classical ground states of the given lattice size with the appropriate operator $\tilde{M}_z$.

\begin{figure*}
    \centering
    \begin{subfigure}[b]{0.3\textwidth}
        \includegraphics[width=54mm, height=39mm]{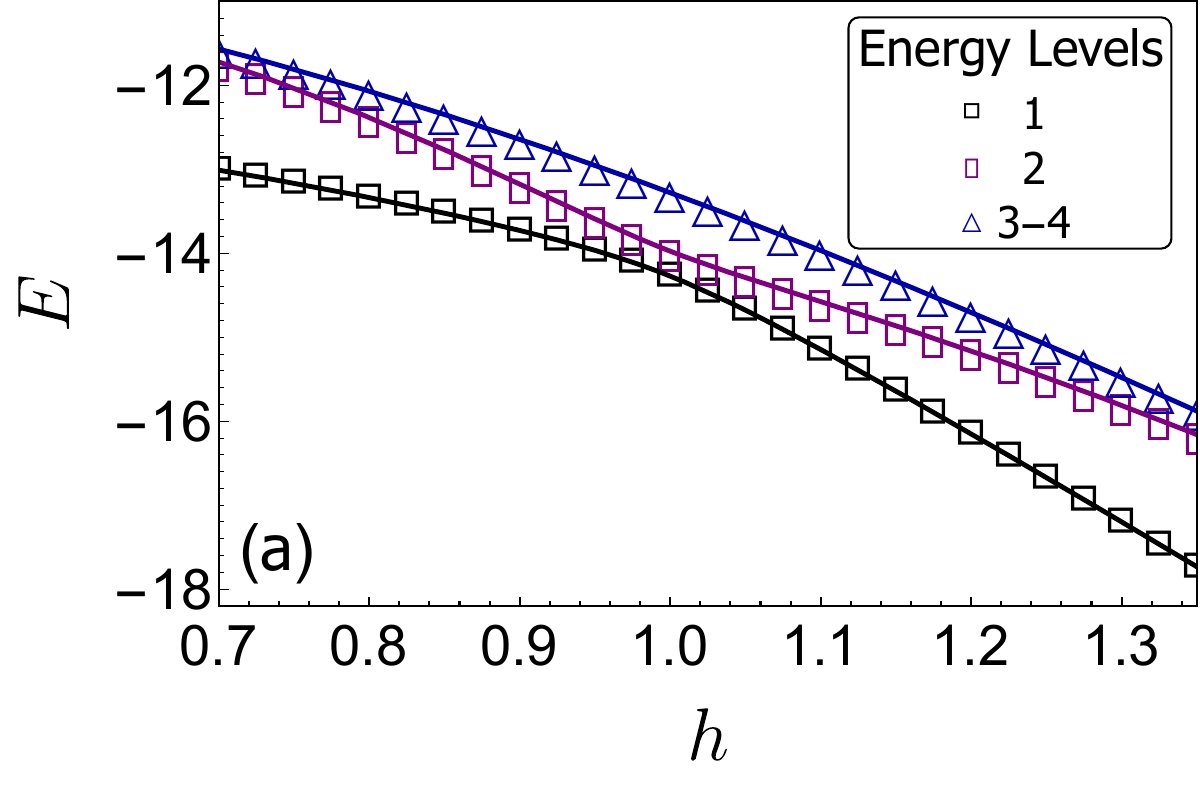}
     \end{subfigure}
     \begin{subfigure}[b]{0.3\textwidth} 
        \includegraphics[width=54mm, height=39mm]{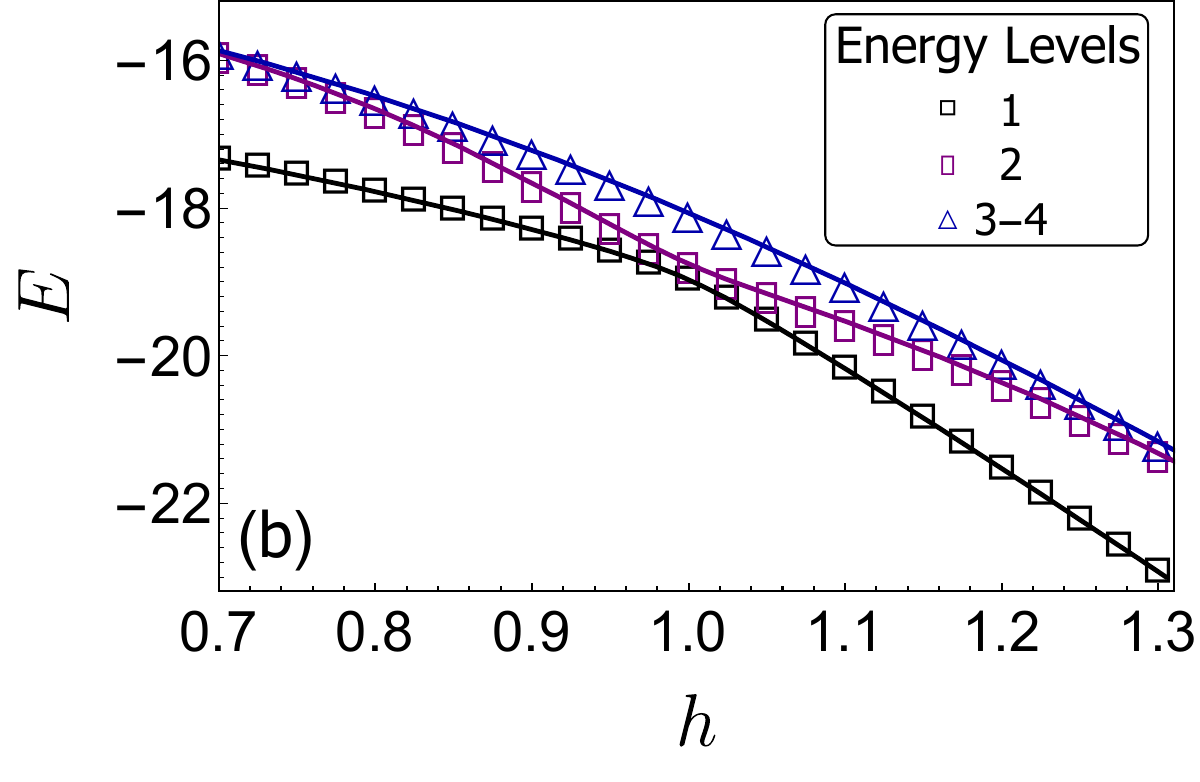}
     \end{subfigure}
     \begin{subfigure}[b]{0.34\textwidth}
        \includegraphics[width=54mm, height=38.5mm]{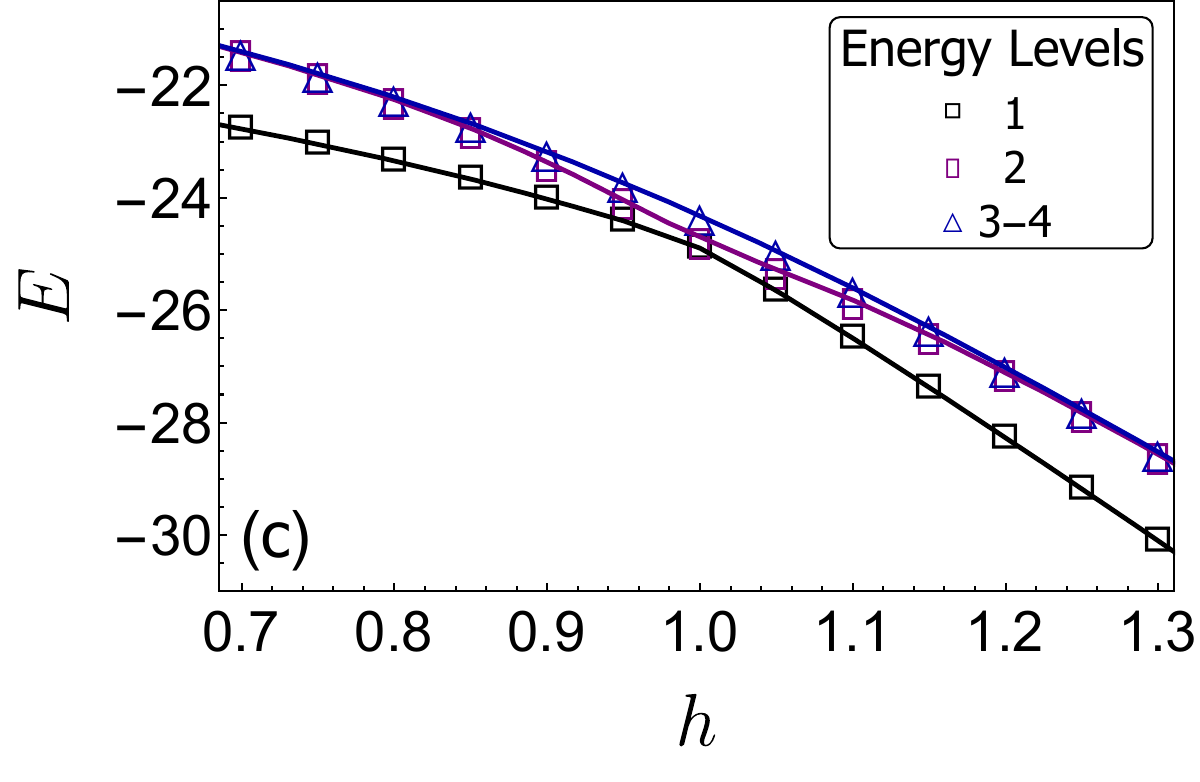}
     \end{subfigure}
     \caption{Low-lying spectrum of the QTPM as a function of $h$ for fixed $J=1$ from ED, for sizes without extra symmetries. (a) $N = 3 \times 4$. (b)  $N = 4 \times 4$. (c)  $N = 3 \times 7$. The avoided crossing between the ground (black squares) and first excited (purple rectangles) states is indicative of a first-order transition.}
     \label{fig:statediagram1st}
\end{figure*}

\begin{figure*}
    \centering
    \begin{subfigure}[b]{0.3\textwidth}
        \includegraphics[width=54mm, height=39mm]{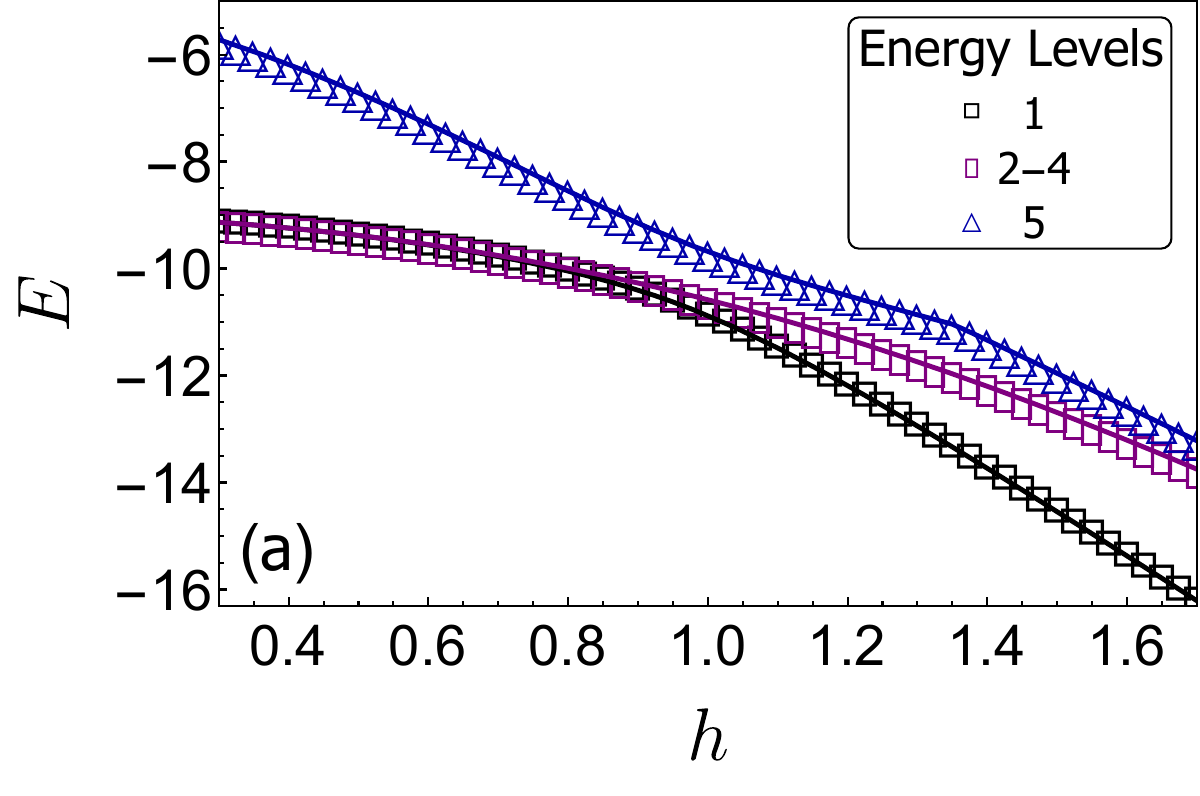}
     \end{subfigure}
     \begin{subfigure}[b]{0.3\textwidth}
        \includegraphics[width=54mm, height=39mm]{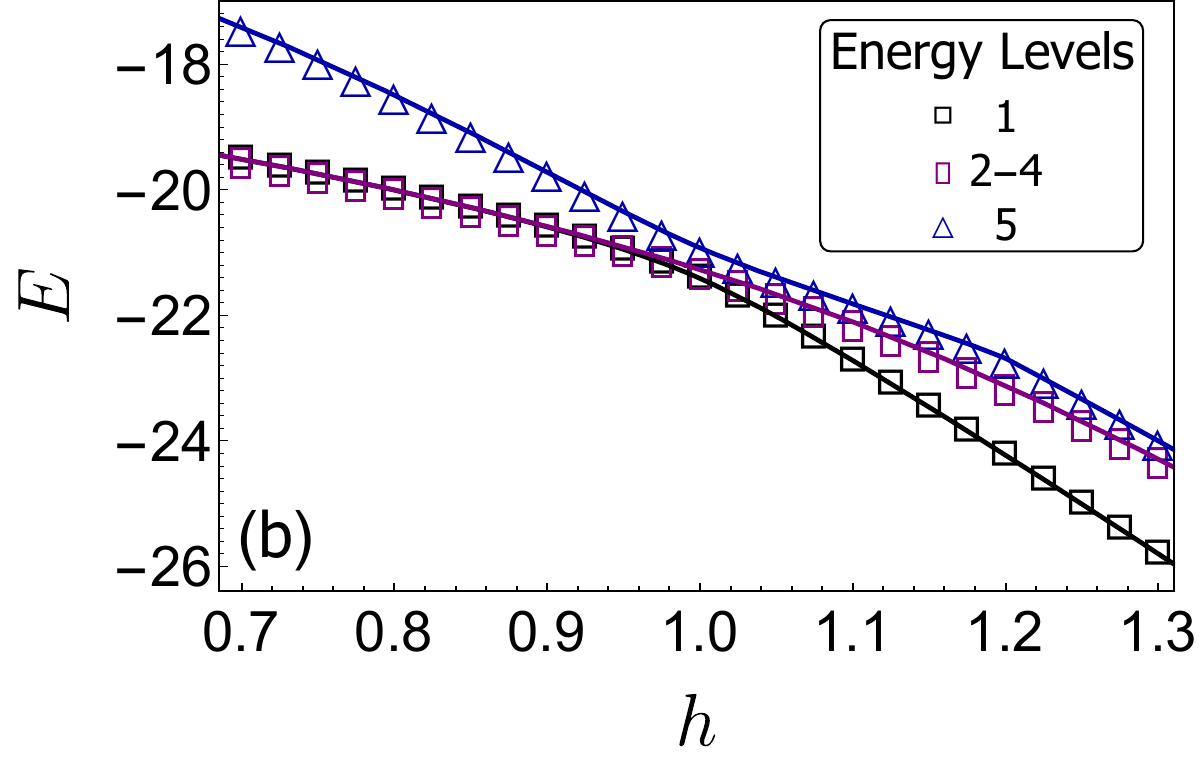}
     \end{subfigure}
     \begin{subfigure}[b]{0.34\textwidth}
        \includegraphics[width=54mm, height=40mm]{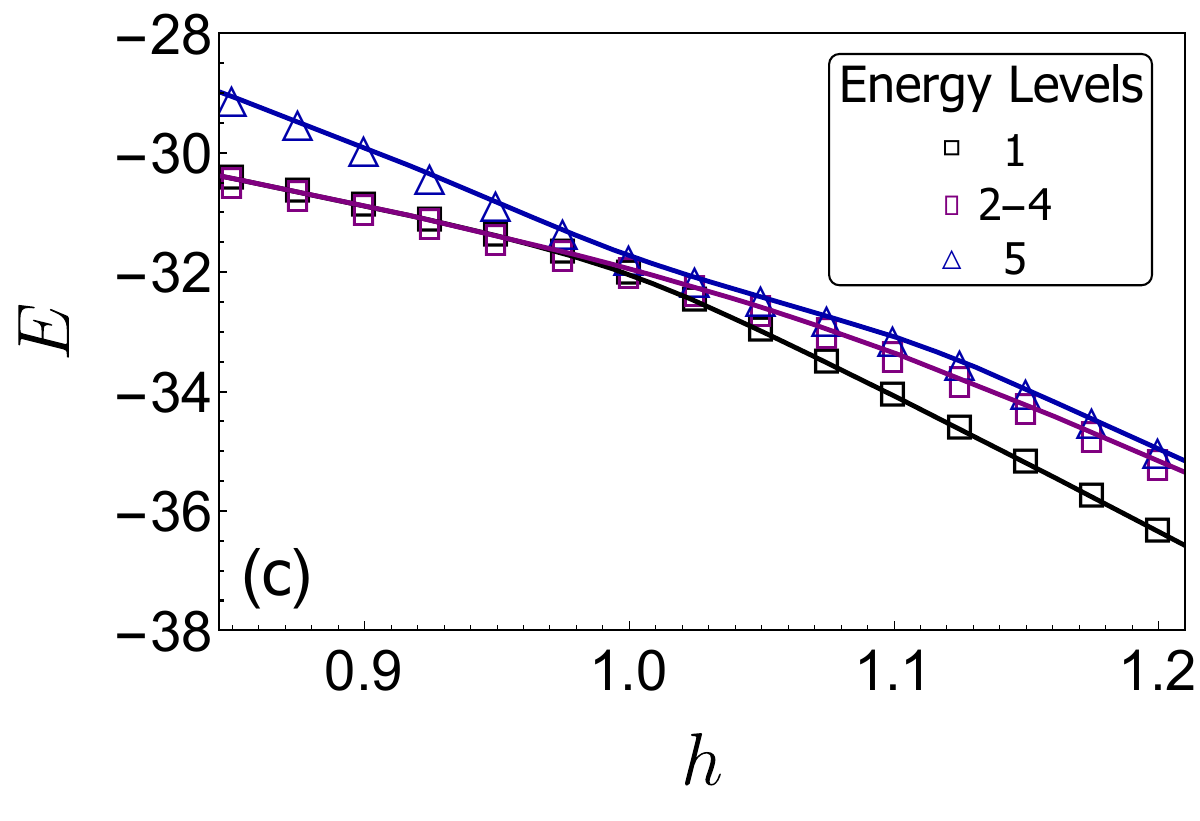}
     \end{subfigure}
     \caption{
     Same as Fig.~\ref{fig:statediagram1st} but for systems with multiple symmetries. (a)  $N = 3 \times 3$. (b) $N = 3 \times 6$. (c) $N = 3 \times 9$. The merging of the ground state (black squares) with three degenerate excited states (purple rectangles) is indicative of spontaneous symmetry breaking. 
     }
     \label{fig:statediagram2nd}
\end{figure*}

In order to more clearly understand the mechanism of the phase transition, in Figs.~\ref{fig:statediagram1st} and \ref{fig:statediagram2nd} we plot the low-lying spectrum of the QTPM from ED as a function of $h$ for fixed $J$. These results support our above observations: for system sizes where only a first-order phase transition is expected, there is an avoided crossing between the ground state and the first excited state; for system sizes with extra symmetries from the cycles of Rule 60, we see {\em both} an avoided crossing (indicative of first-order transitions) {\em and} a merging of eigenstates indicative of spontaneous symmetry breaking. As seen in Fig.~\ref{fig:statediagram2nd} for the case of $N = 3 \times M$, the avoided crossing becomes apparent only with increasing system size. 

We now comment on how our results compare to those in Ref.~\cite{2021_Zhou_Pollmann}. For the numerics, Ref.~\cite{2021_Zhou_Pollmann} used a stochastic series expansion (SSE) approach. We in turn use MPS and ctQMC. Both SSE and ctQMC are Quantum Monte Carlo based methods, which indicates that they, in principle, should be able to roughly access system sizes of the same order of magnitude.

Furthermore, while Ref.~\cite{2021_Zhou_Pollmann} also considered PBCs, there was no specific restriction on system size, and therefore no distinction between sizes for which there is a single classical minimum and sizes where there are multiple ones, with the implications for symmetries of the corresponding QTPM. Ref.~\cite{2021_Zhou_Pollmann} also used a non-local order parameter, compared to our local ones (the staggered magnetisations) that do reflect the minima of the underlying TPM. In \cite{2021_Zhou_Pollmann}, the existence of a phase transition at $J=h$ was confirmed through the study of the Binder cumulant; this was done, however, with limited accuracy on the location of the phase transition point. It is important to note that some of the local observables we calculate here are also studied for specific system sizes in the Appendix of Ref.~\cite{2021_Zhou_Pollmann}. Since the temperature used for those calculations varied for different system sizes, it is possible that the smoothness observed in Ref.~\cite{2021_Zhou_Pollmann} is a consequence of thermal effects. We instead used a {\em fixed} inverse temperature $\beta=128$ which we verified is sufficient to make thermal effects negligible.

\subsection{Nature of the phase transition in the thermodynamic limit}

The discussion above and the numerical results indicate the existence of a quantum phase transition in the thermodynamic limit, $N \to \infty$, at the self-dual point, $J = h$, of the QTPM. However, the approach to the thermodynamic limit is different across different system size geometries.

There are three different limits to thermodynamics: (i) across one of the two dimensions while the other one remains fixed (that is, infinite strips), (ii) across both dimensions (e.g. square sizes), and (iii) on making the spins continuous. We briefly discuss the differences between these limits and the complications that might arise.

In the case (i), if the limit is taken for fixed $L$ and with $M$ such that $\lcm{\mathcal{C}} | M$ (e.g. $M=3k$, with $k  \in \mathbb{N}$), the number of classical ground states remains the same. In our numerics we are restricted to narrow strips to allow convergence of the MPS algorithm. Fig.~\ref{fig:MPS_ED_ctQMC_rectangular} suggests that in such quasi-1D systems the transition will eventually be slighly weaker than for square system sizes. 

Case (ii) can be more involved. The simplest situation is that of square lattices $N = L \times L$ with $L$ a power of two, where it is guaranteed that for all sizes there will be a single classical ground state, and therefore the transition is certainly first-order. For other size sequences,  the number of relevant Rule 60 cycles, and therefore symmetries of the QTPM, may grow or decline with system size. For some cases this growth is monotonic (as for example for $N = 3^k\times 3^k$ with $k\rightarrow \infty$), while in others it is not (as for example when $N = 3k\times 3k$ with $k\rightarrow \infty$), see Table \ref{fig:Cycle_Lengths_up_to_40}.

In case (iii) the nature of the underlying CA is altered \cite{2000_Flocchini,2005_Mingarelli}. In this limit, Rule 60 becomes
\begin{equation}
    s = f(p, q, r) = p+q - 2pq,
\end{equation}
where $p$, $q$ and $r$ indicate the state of the three sites in the neighbourhood of site $s$, determining the local evolution of the CA, see Fig.~\ref{fig:Ruleplots}. Basic arguments \cite{2000_Flocchini} indicate a single fixed point in the evolution of this \textit{fuzzy} CA. We speculate that the same behaviour will be observed in the quantum field theory limit for QTPM; a single ground state across different regions of the whole $J-h$ space and thus a first-order quantum phase transition. However, a field theoretic description of the QTPM might not be as obvious and straightforward to get for the above elementary argument to hold.

\section{Conclusions}{\label{conclusions}}

In this work, we used the cycles of the cellular automaton Rule 60 to describe the symmetries of the quantum triangular plaquette model. We found that the attractor structure of Rule 60 plays an important role in the characterisation of the degeneracies of the ground states of the classical TPM, allowing in turn the construction od the symmetry operators of the QTPM. In this way, the existence or absence of stable cycles for Rule 60 imply whether it is possible or not for the QTPM to display SSB, which in turn impacts the nature of the quantum phase transition at the self-dual point. These general observations are also consistent with the finite size trends from our numerical simulations. 
    In contrast to our work here, 
    Ref.~\cite{2021_Zhou_Pollmann} considered square lattices with PBC but without distinguishing systems with multiple or single classical ground states. Our results here indicate that systems with single classical minima (as those in Ref.~\cite{2021_Zhou_Pollmann}) always have a first-order quantum phase transitions, with the addition of SSB in the low lying spectra for other system sizes.

A full description of the QTPM phase transition would require a field theoretical description and a renormalization group treatment; we leave these tasks for future works.
    Another possible extension of our work, for example using the method of Ref.~\cite{2023_Causer}, is the analysis of the finite temperature phase diagram of quantum TPM. As implied by Refs.~\cite{2020_Grover,2020_Grover_2,2023_boulder_school_grover}, for the $T\neq0$ case, when there does not exist an extensive number of constraints of products of spins equal to one, then there exists no thermal phase transition. The reverse is not always true. Most cases of models with an extensive number of constraints of products of spins equal to one exhibit a finite temperature phase transition, which is what we also expect for the quantum TPM

Lastly, for system sizes where multiple ground states are encountered we observe the energy gap vanishing exponentially fast, see Appendix \ref{appendixB}. In contrast, for the 3-XORSAT instance studied in Ref.~\cite{2021_Medina}, the gap vanishes only polynomially. Further studies would be needed for resolving the origin of this discrepancy, as it would reflect on the applicability of quantum annealing. 

\begin{acknowledgments}
We thank 
S. Balasubramanian, N. Calkin, J.C\^{o}t\'{e}, A. Fahimniya, E. Hecker, T. Grover, J. Lahtonen, E.Lake, C. Li, A. Smith, N. Tandivasadakarn, M. Tikhanovskaya, L. Vasiloiu and M. Will for discussions.

We acknowledge financial support from EPSRC Grant no.\ EP/R04421X/1, the Leverhulme Trust Grant No. RPG-2018-181, and University of Nottingham grant no.\ FiF1/3.
LC was supported by an EPSRC Doctoral prize from the University of Nottingham.
Simulations were performed using the University of Nottingham Augusta HPC cluster, and the Sulis Tier 2 HPC platform hosted by the Scientific Computing Research Technology Platform at the University of Warwick (funded by EPSRC Grant EP/T022108/1 and the HPC Midlands+ consortium). 
\end{acknowledgments}

\bibliography{references_trial}

\appendix
\section{TPM and QTPM for other boundary conditions}{\label{appendixA}}

In the main text, we show that the ground state properties of the TPM and, consequently, the quantum phase transition of the QTPM depend on the boundary conditions, but we only focused on systems with PBC. Here, we consider the cases of periodic boundaries in only the x-dimension (PBCx) and of open boundaries (OBC), using the same Rule 60 CA considerations as for the periodic case.

For PBCx, we use the update rule for Rule 60 as before, with the only difference that we do not need to explicitly check for the periodicity across the $y$-direction. As a result, for an initial array of $L$ sites, there will be $2^L$ configurations and, thus, $2^L$ ground states for the classical TPM. In this case, only the number of sites in the $x$-direction matters for the number of classical ground states. For example, a lattice with size $N = 3 \times 3$ and one with $N = 3 \times 80$ will have the same number, 8, of ground states. The identification of the classical ground states can be worked out from the Rule 60 evolution, as before.

For OBC, the update rule for Rule 60 is modified for the first cell of an L-length array so that it is not updated. This freedom on choosing two of the boundaries of the lattice gives an increased number of ground states for the classical TPM. Specifically, given a lattice of N spins, $N = L \times M$, the number of the classical ground states is $2^{L + M - 1}$. In both cases the number of classical ground states grows unboundedly, but subextensively, with the system size.

We now perform a similar numerical analysis as in Sec.~\ref{numerics}, but only using MPS methods. Data is normalised with the system size of the given lattice. The system sizes accessible do not give a clear indication of a well-formed phase transition, but only signatures of it. The first-order transition found is weaker than in the case with fully PBC, which we attribute to the high number of ground states for the classical TPM, given the system sizes. We note that all these states for $h \neq 0$ constitute low-lying excited states which affect the convergence of the MPS algorithm and hinder the signature of the avoided gap crossing.

\begin{figure*}
    \centering
    \begin{subfigure}[b]{0.3\textwidth}
        \includegraphics[width=50mm, height=35mm]{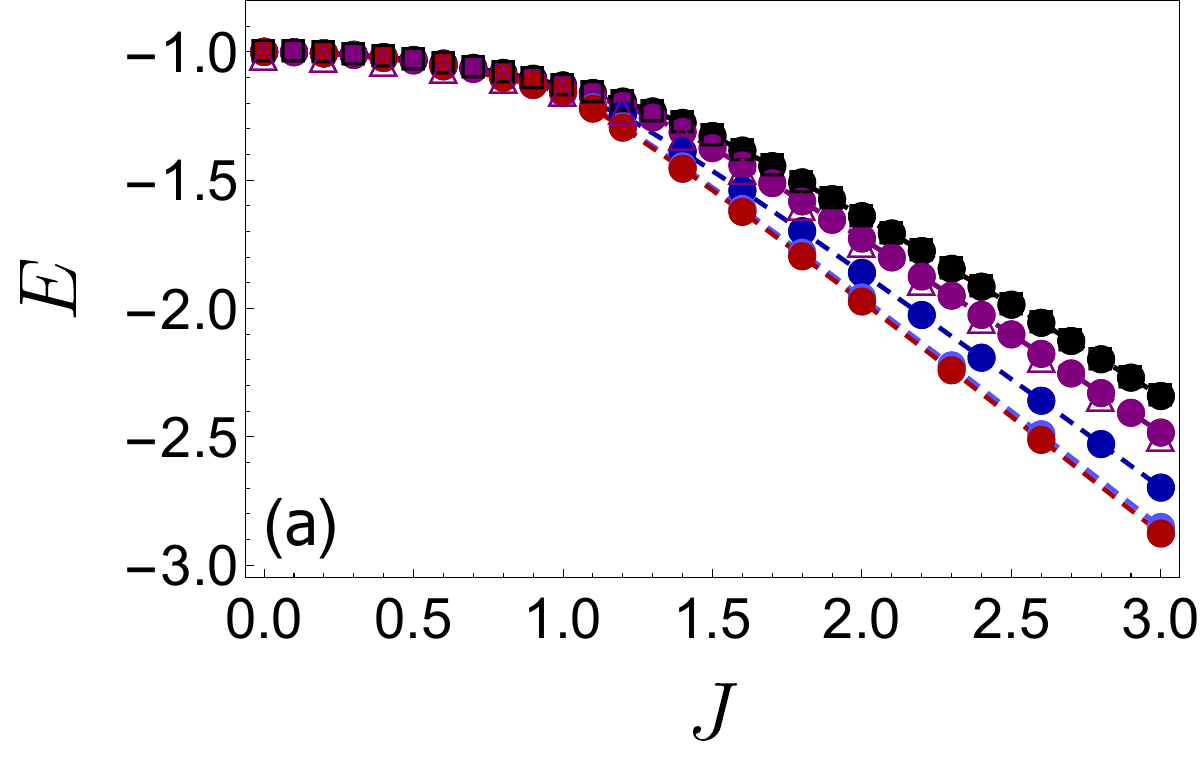}
     \end{subfigure}
     \begin{subfigure}[b]{0.3\textwidth}
        \includegraphics[width=50mm, height=35mm]{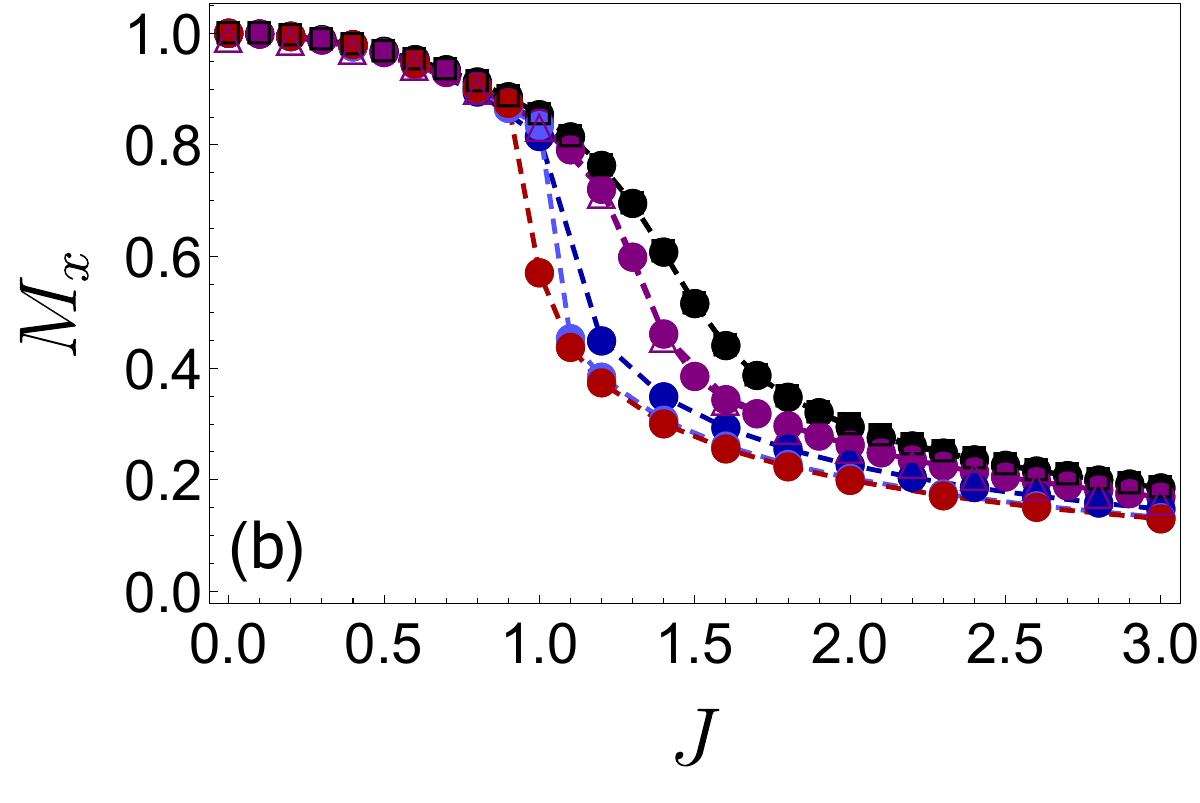}
     \end{subfigure}
     \begin{subfigure}[b]{0.3\textwidth}
        \includegraphics[width=50mm, height=35mm]{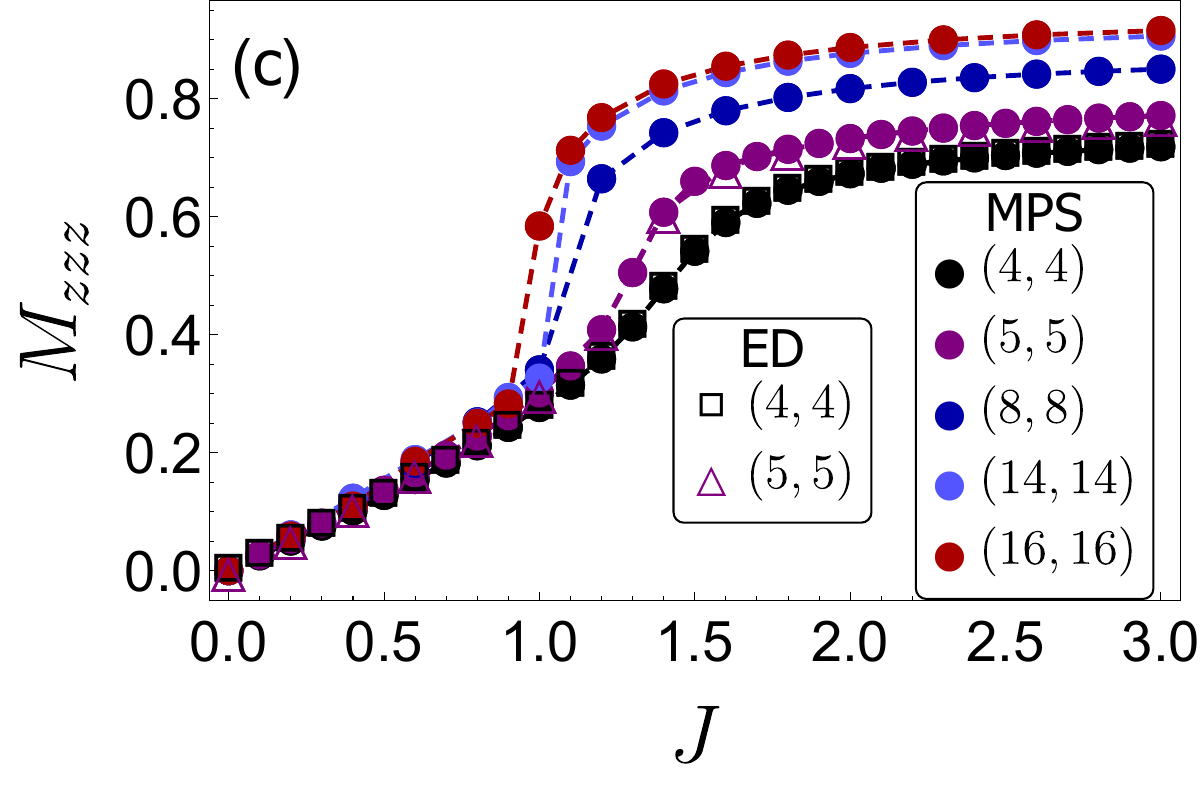}
     \end{subfigure}
     \caption[PBCx square]{(a) Normalised energy, (b) $M_x$ and (c) $M_{zzz}$ of ground states from MPS and ED for
     square lattices with PBCx. Data from ED are denoted as empty squares and empty triangles.}
     \label{fig:MPS_ED_PBCx_square}
\end{figure*}

\begin{figure*}
    \centering
    \begin{subfigure}[b]{0.3\textwidth}
        \includegraphics[width=50mm, height=35mm]{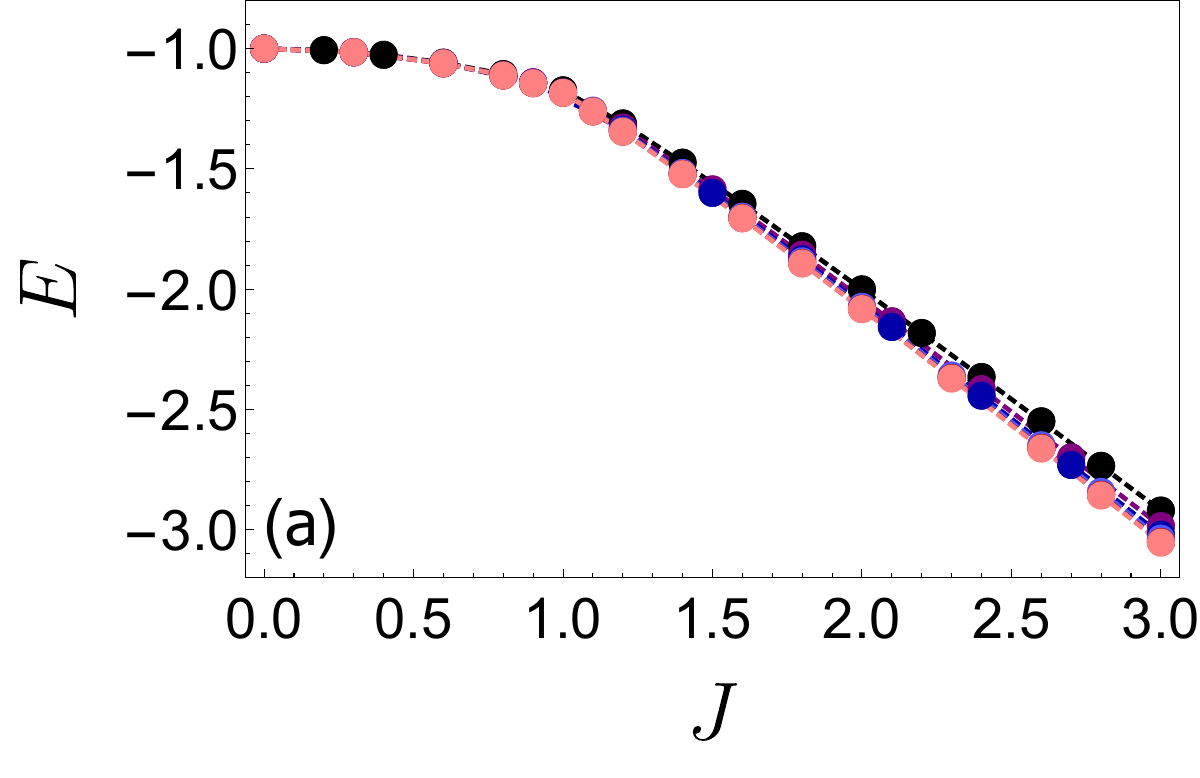}
     \end{subfigure}
     \begin{subfigure}[b]{0.3\textwidth}
        \includegraphics[width=50mm, height=35mm]{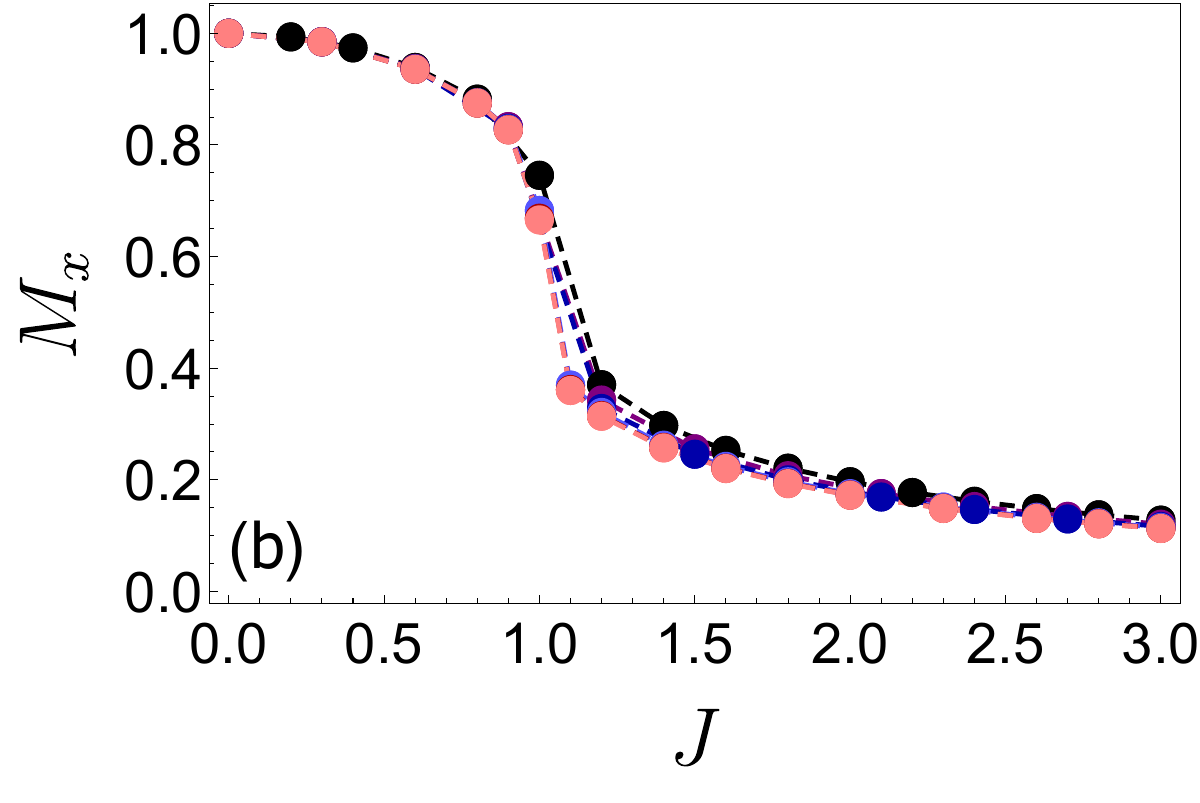}
     \end{subfigure}
     \begin{subfigure}[b]{0.3\textwidth}
        \includegraphics[width=50mm, height=35mm]{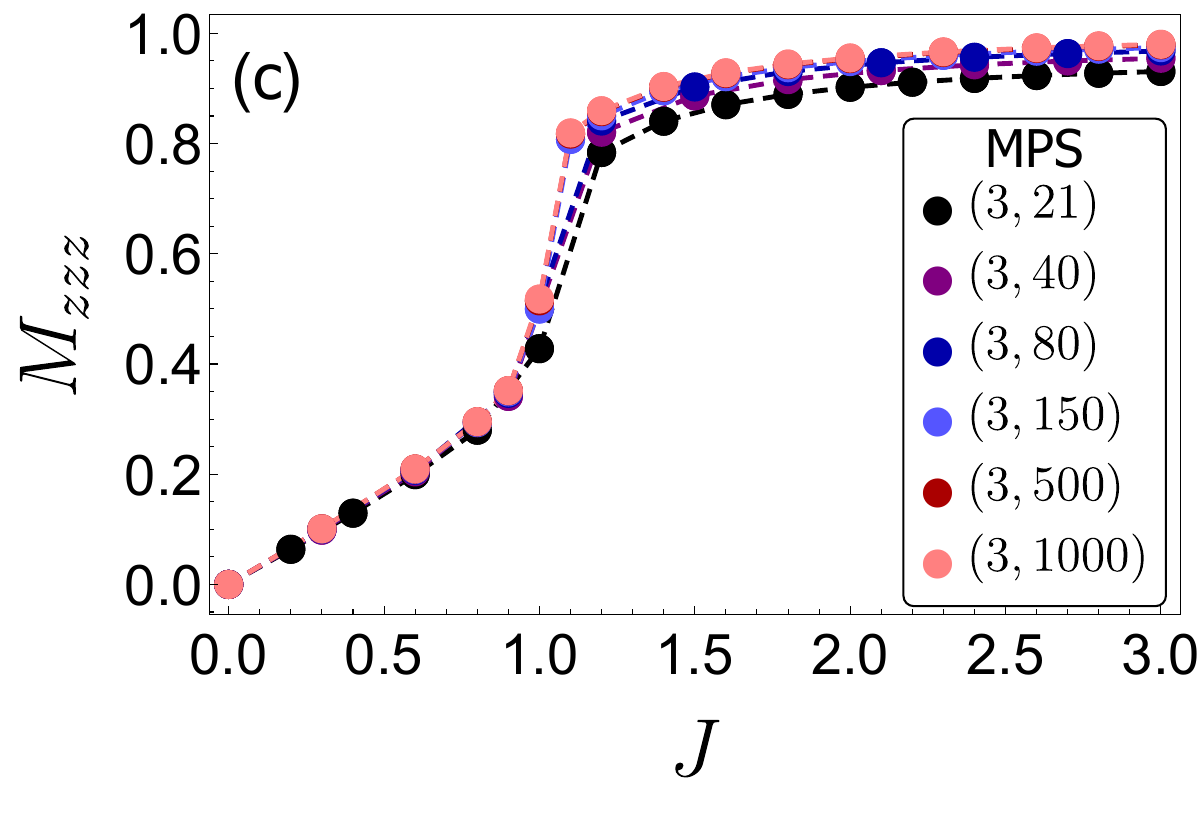}
     \end{subfigure}
     \caption[PBCx rectangular]{Same as Fig.~\ref{fig:MPS_ED_PBCx_square} but for systems of size $N = 3 \times L$.}
     \label{fig:MPS_ED_PBCx_rectangular}
\end{figure*}

As seen from Figs.~\ref{fig:MPS_ED_PBCx_square} and \ref{fig:MPS_ED_PBCx_rectangular} for PBCx, the difference between the square lattice size scaling and the quasi-1D rectangular strips is more pronounced, when compared to the finite-size scaling for PBC. Extra calculations on wider rectangular strips verify that this difference is only a feature of the quasi-1D geometry of the lattice and not an inherent property of the system. Accuracy is lost with increasing size and the MPS results for the sizes studied are not reflective of the true thermodynamic limit. 

\begin{figure*}
    \centering
    \begin{subfigure}[b]{0.3\textwidth}
        \includegraphics[width=50mm, height=35mm]{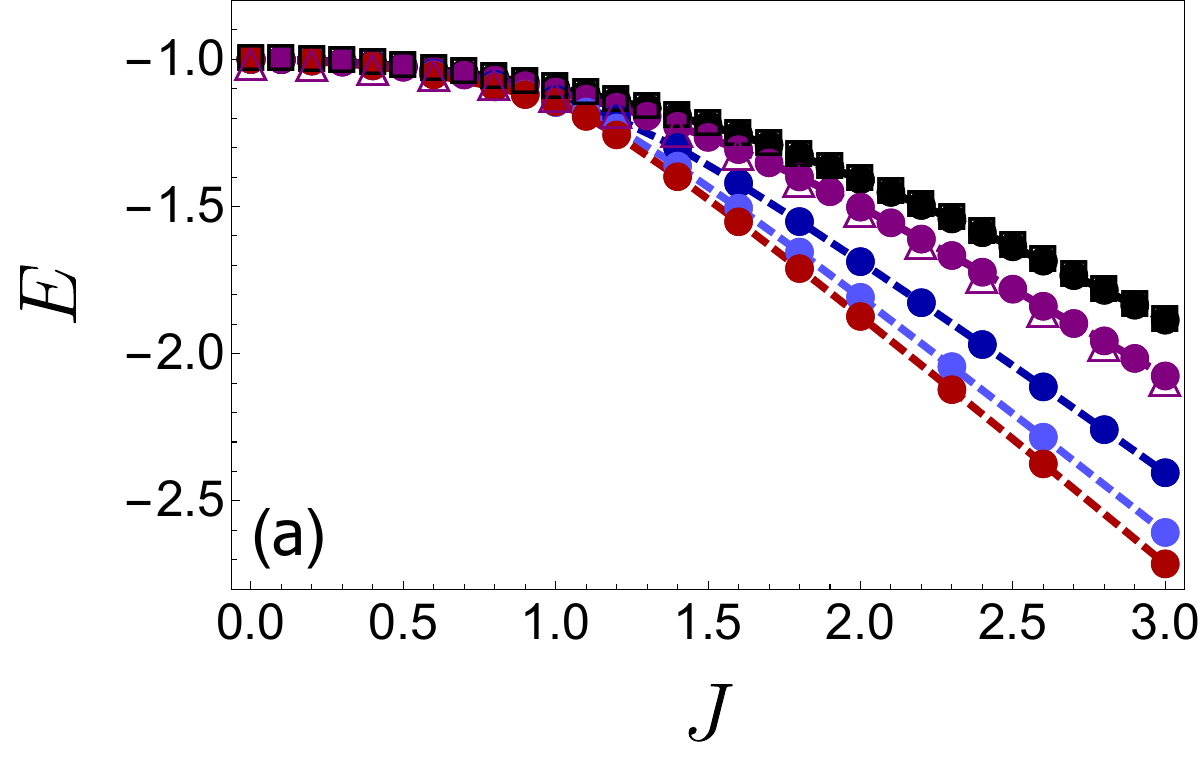}
     \end{subfigure}
     \begin{subfigure}[b]{0.3\textwidth}
        \includegraphics[width=50mm, height=35mm]{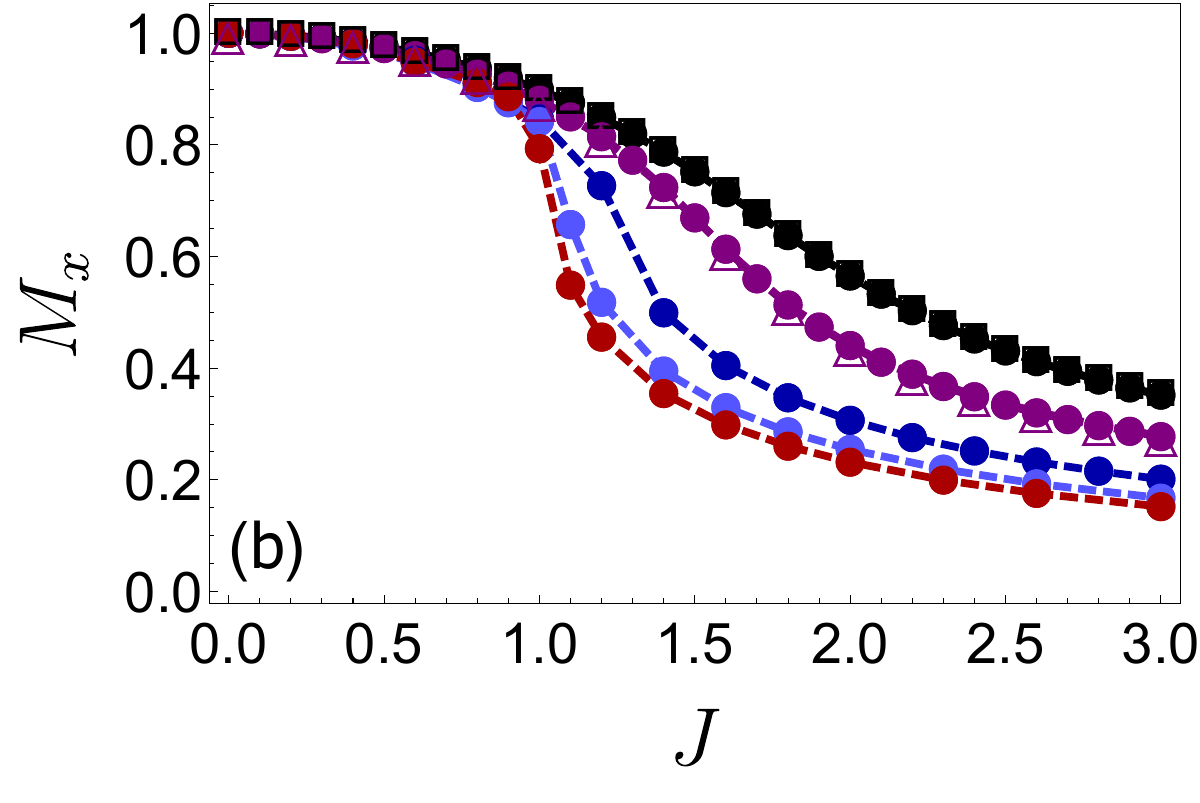}
     \end{subfigure}
     \begin{subfigure}[b]{0.3\textwidth}
        \includegraphics[width=50mm, height=35mm]{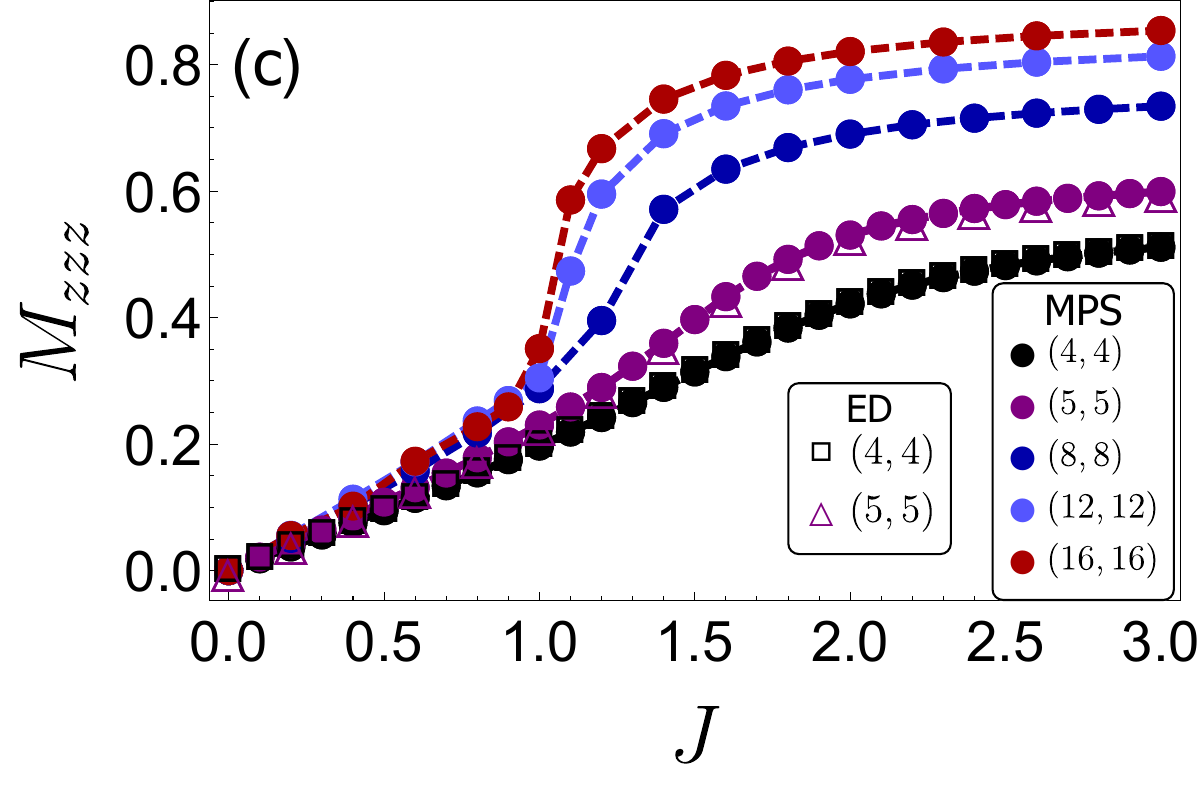}
     \end{subfigure}
     \caption[OBC square]{(a) Normalised energy, (b) $M_x$ and (c) $M_{zzz}$ of ground states from MPS and ED for
     square lattices with OBC. Data from ED are denoted as empty squares and empty triangles.}
     \label{fig:MPS_ED_OBC_square}
\end{figure*}

\begin{figure*}
    \centering
    \begin{subfigure}[b]{0.3\textwidth}
        \includegraphics[width=50mm, height=35mm]{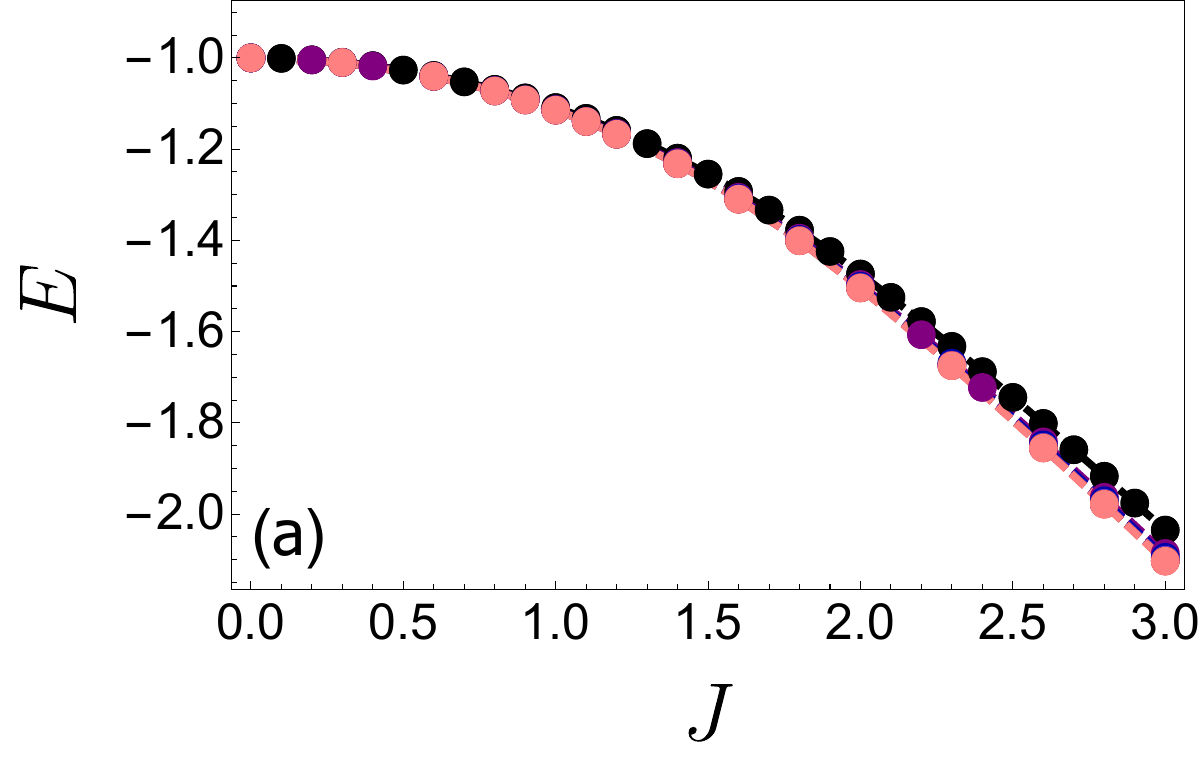}
     \end{subfigure}
     \begin{subfigure}[b]{0.3\textwidth}
        \includegraphics[width=50mm, height=35mm]{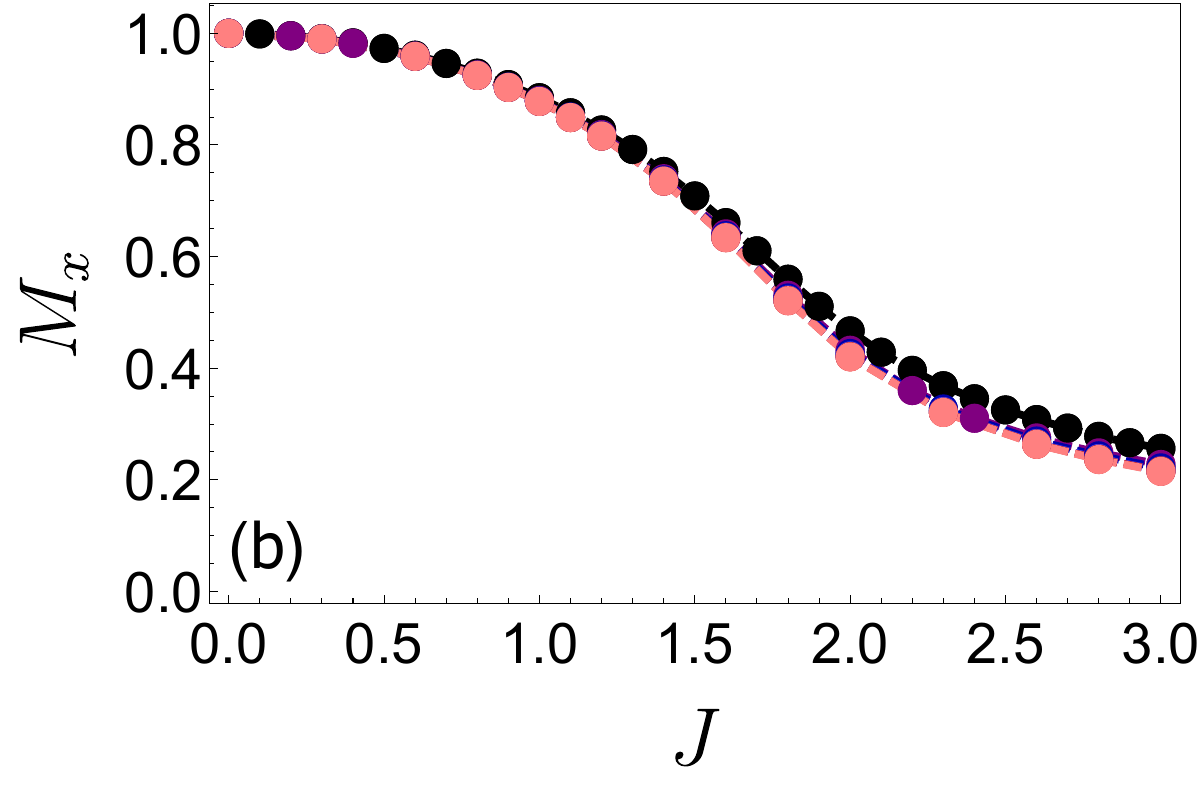}
     \end{subfigure}
     \begin{subfigure}[b]{0.3\textwidth}
        \includegraphics[width=50mm, height=35mm]{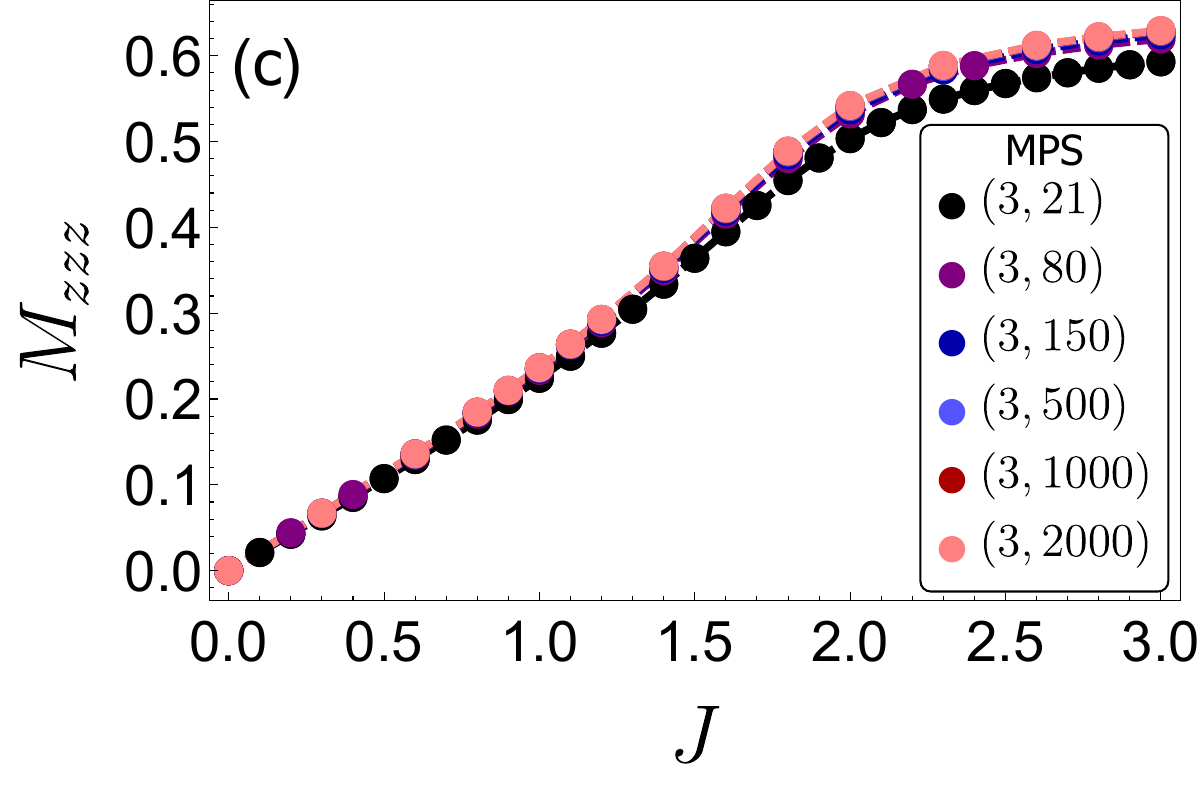}
     \end{subfigure}
     \caption[OBC rectangular]{Same as Fig.~\ref{fig:MPS_ED_OBC_square} but for systems of size $N = 3 \times L$.}
     \label{fig:MPS_ED_OBC_rectangular}
\end{figure*}

The above considerations for PBCx are even more noticeable for the case of OBC, compare Figs.~\ref{fig:MPS_ED_OBC_square} and \ref{fig:MPS_ED_OBC_rectangular}. For the quasi-1D strips we see smooth behaviour for all values of $J$, while they seem to have converged to their ``thermodynamic'' behaviour. However, as seen from the square system sizes, the behaviour of the model remains the same regardless of the boundary conditions. It becomes apparent though that bigger system sizes soon become computationally inaccessible due to the exponential number of classical ground states. This behaviour shows an obvious discrepancy with standard MPS methods; normally, fully periodic system sizes are computationally harder to access. Here, since the model 
can have an exponential number of low-lying states(ground states for $h=0$) for OBC, the convergence of the algorithm is hindered and significantly increases the lattice size where the ``thermodynamic limit'' has been reached. Therefore, only for PBC, the thermodynamic limit becomes apparent for the sizes we can access.

\begin{figure*}
    \centering
    \begin{subfigure}[b]{0.45\textwidth}
        \includegraphics[width=8cm, height=5cm]{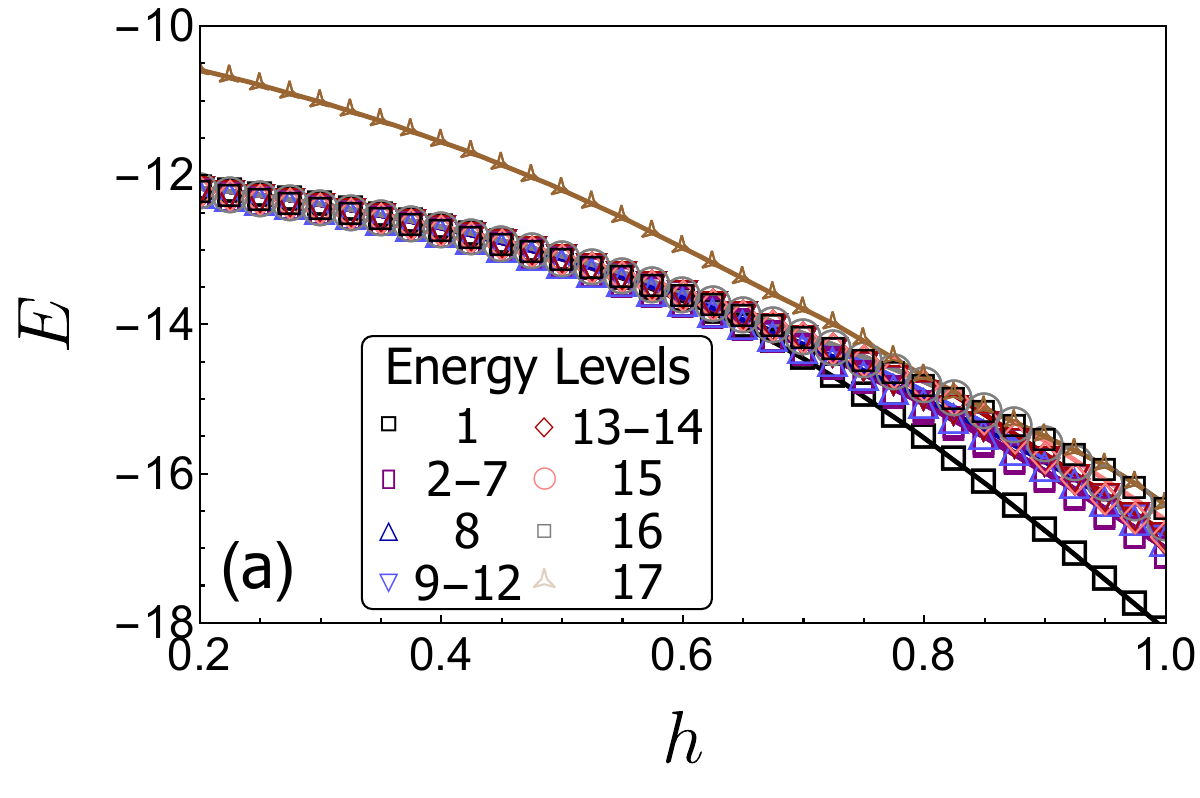}
    \end{subfigure}
    \begin{subfigure}[b]{0.45\textwidth}
        \includegraphics[width=8cm, height=5cm]{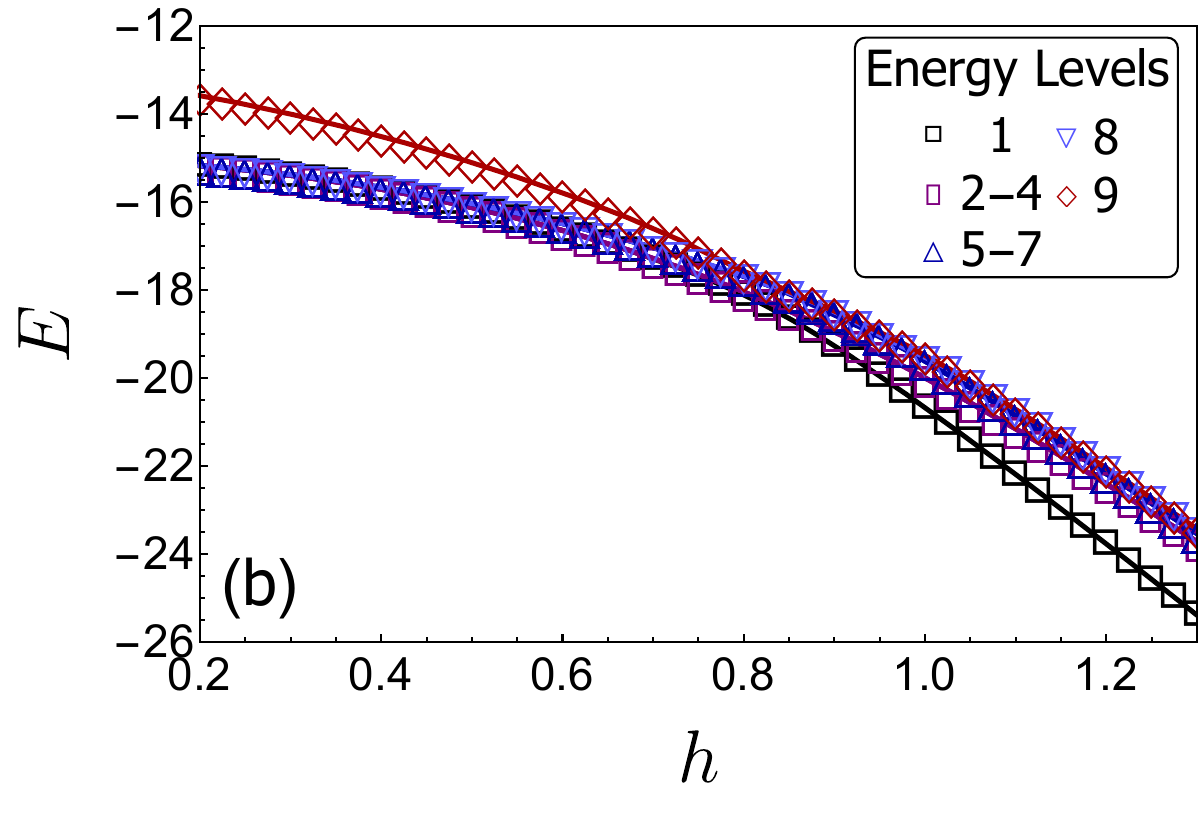}
    \end{subfigure}
    \caption{\label{fig:statediagramPBCx} The unnormalised state diagrams for a (a) $4 \times 4$ and a (b) $3 \times 6$ lattice with PBCx.}
\end{figure*}

\begin{figure*}
    \centering
    \begin{subfigure}[b]{0.45\textwidth}
        \includegraphics[width=8cm, height=5cm]{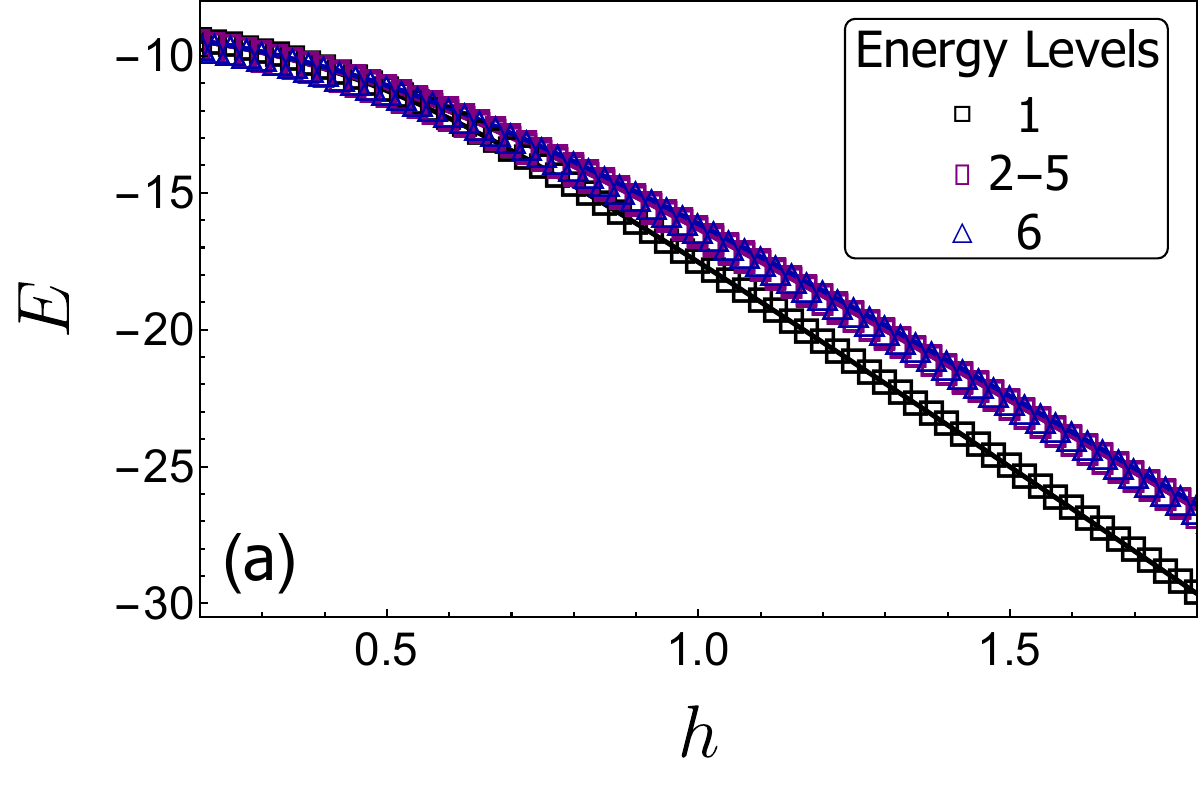}
    \end{subfigure}
    \begin{subfigure}[b]{0.45\textwidth}
        \includegraphics[width=8cm, height=5cm]{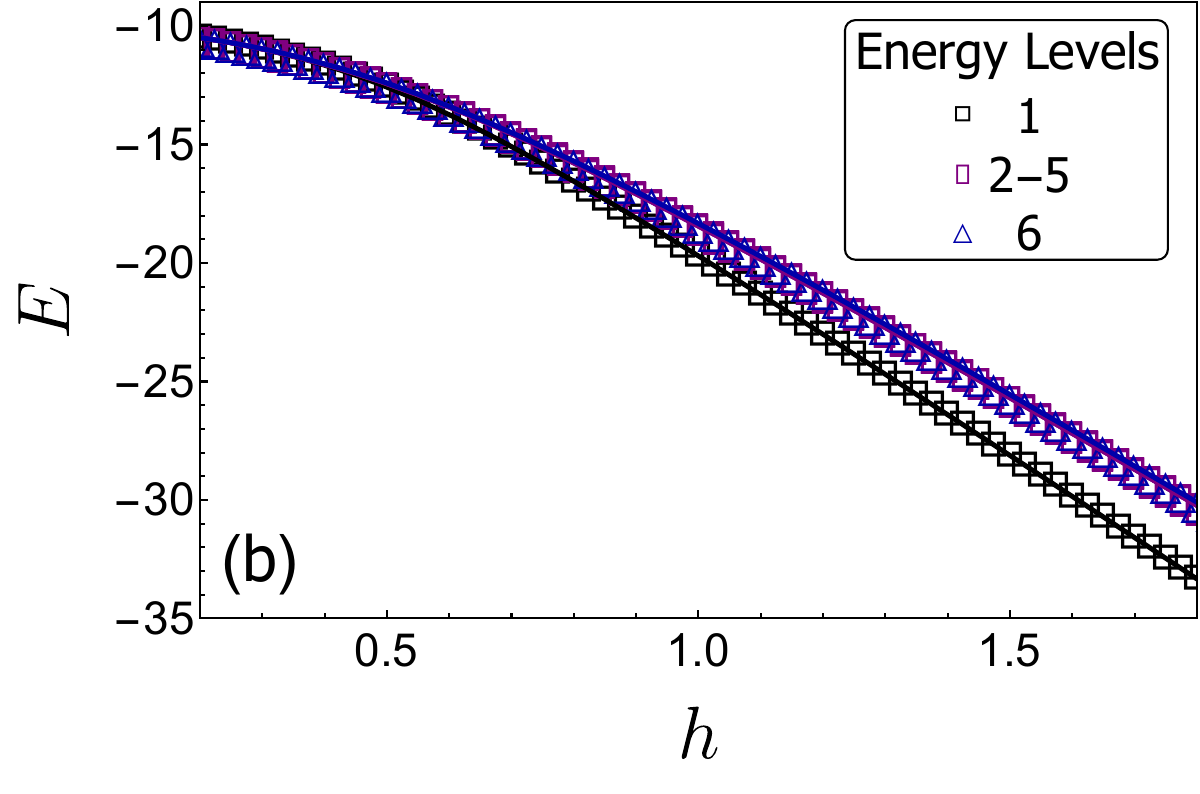}
    \end{subfigure}
    \caption{\label{fig:statediagramOBC} Same as Fig.~\ref{fig:statediagramPBCx} for OBC.}
\end{figure*}

The significance of these arguments is further evident from Figs.~\ref{fig:statediagramPBCx} and \ref{fig:statediagramOBC}. For the case of PBCx, all degenerate ground states for the $J \gg h$ region are easily found from exact diagonalization calculations and classically excited states are easily tractable too. However, the same is not true for OBC. The number of classically degenerate ground states increases exponentially and this is the reason why it would be pointless to show more ground states. The number of low-lying excited states which form the classical ground state degeneracy are also the ones to blaim for obscuring of the quantum phase transition in the later case; the avoided gap crossing occurs between the ground state and the ($\cal M $ $+ 1 $)-th excited state, which shows a reduced (or even nonexistent) signature on the discontinuity of the quantum phase transition, hidden by the massive SSB of the ground state.

\section{Gap Scaling Analysis for PBC}{\label{appendixB}}

In this section we present a restricted and with limited accuracy analysis on the energy difference between the ground state and the first excited state. This analysis was conducted based on ED and MPS methods, which limits the validity of the conclusions that can be reached: it becomes quickly obvious that MPS methods are not powerful enough for the detection of the actual gap, especially in regions of the parameter space with high entanglement or with a high number of low-lying excited states, where MPS often converge to excited states above the lowest-lying ones. However, the analysis below still provides an indication of the behaviour of the gap with system size when comparing systems with different symmetries. 

\begin{figure*}
    \centering
    \begin{subfigure}[b]{0.45\textwidth}
        \includegraphics[width=8cm, height=5cm]{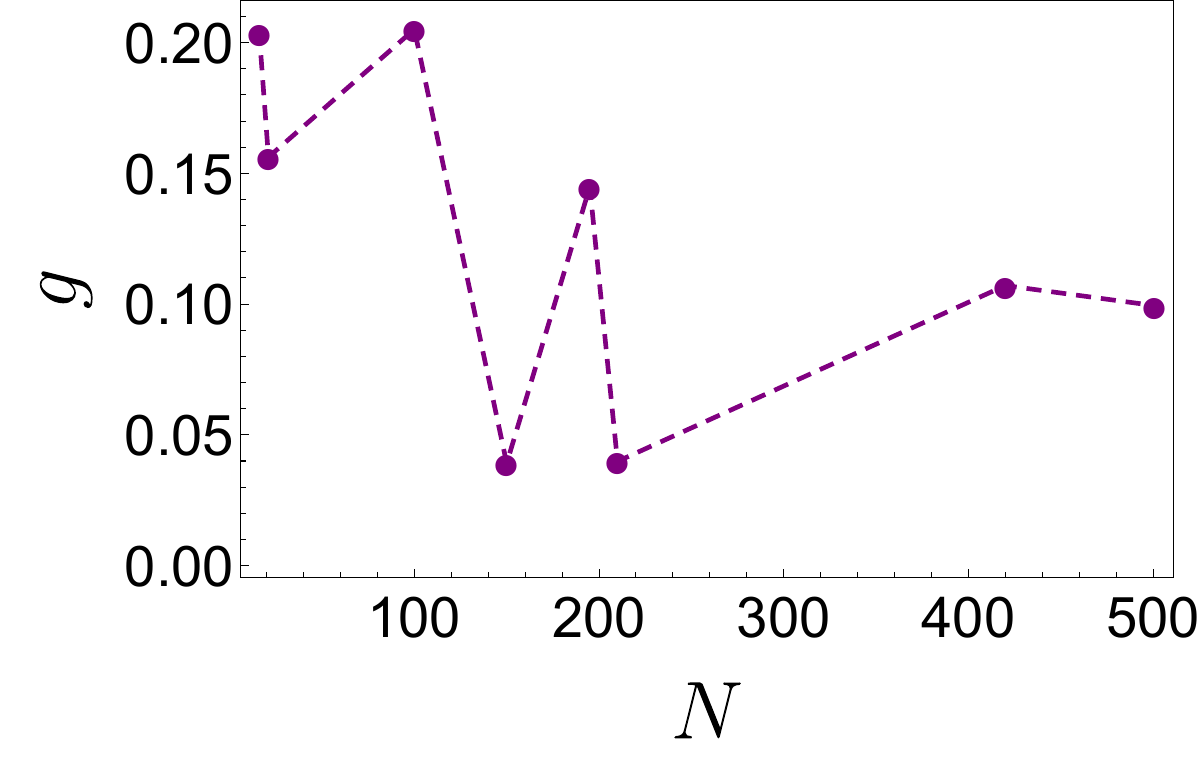}
    \end{subfigure}
    \begin{subfigure}[b]{0.45\textwidth}
        \includegraphics[width=8cm, height=5cm]{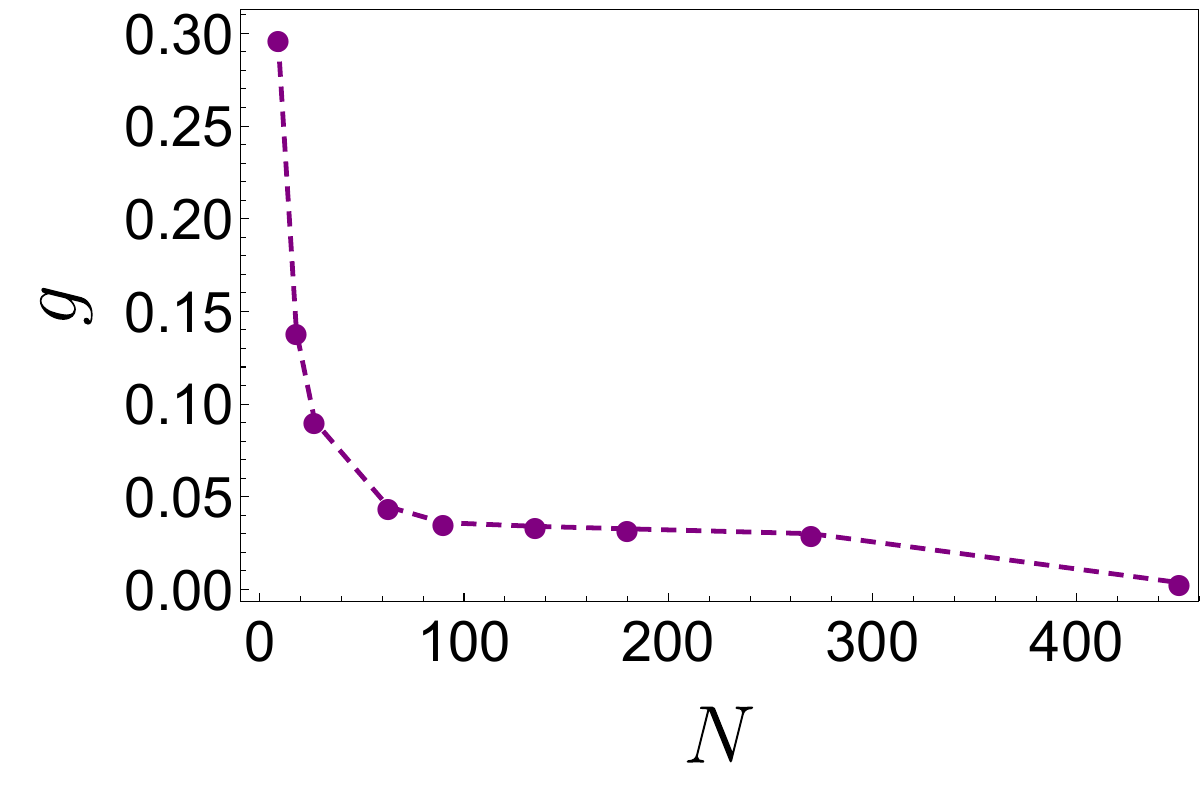}  
    \end{subfigure}
    \caption{
        \label{fig:gapscaling} 
        The scaling of the gap, $g$, for different lattice sizes without (a) and with (b) symmetries for the QTPM for the $J=h=1.0$ point. 
        Both ED and MPS methods are used (where appropriate) for the calculation of the given gaps. Square and rectangular sizes are equally used.}
\end{figure*}

This limited accuracy when measuring the first excited state energy is evident in Fig.~\ref{fig:gapscaling}(a). Data is calculated for the $J=h=1.0$ point. The gap seems to decrease with increasing the system size, but at the same time, the power of MPS to detect it is significantly reduced. The situation seems clearer for Fig.~\ref{fig:gapscaling}(b). However, it is equally problematic despite the monotonically decreasing gap. The only significance of these results are as upper bounds of the actual gap. For the first case, the gap seems to decrease algebraically to zero, while for the case of multiple classical ground states, it seems to decrease exponentially. This underlines the different behaviour depending on the existence or not of multiple classical ground states.  

\begin{figure*}
    \centering
    \begin{subfigure}[b]{0.45\textwidth}
        \includegraphics[width=8cm, height=5cm]{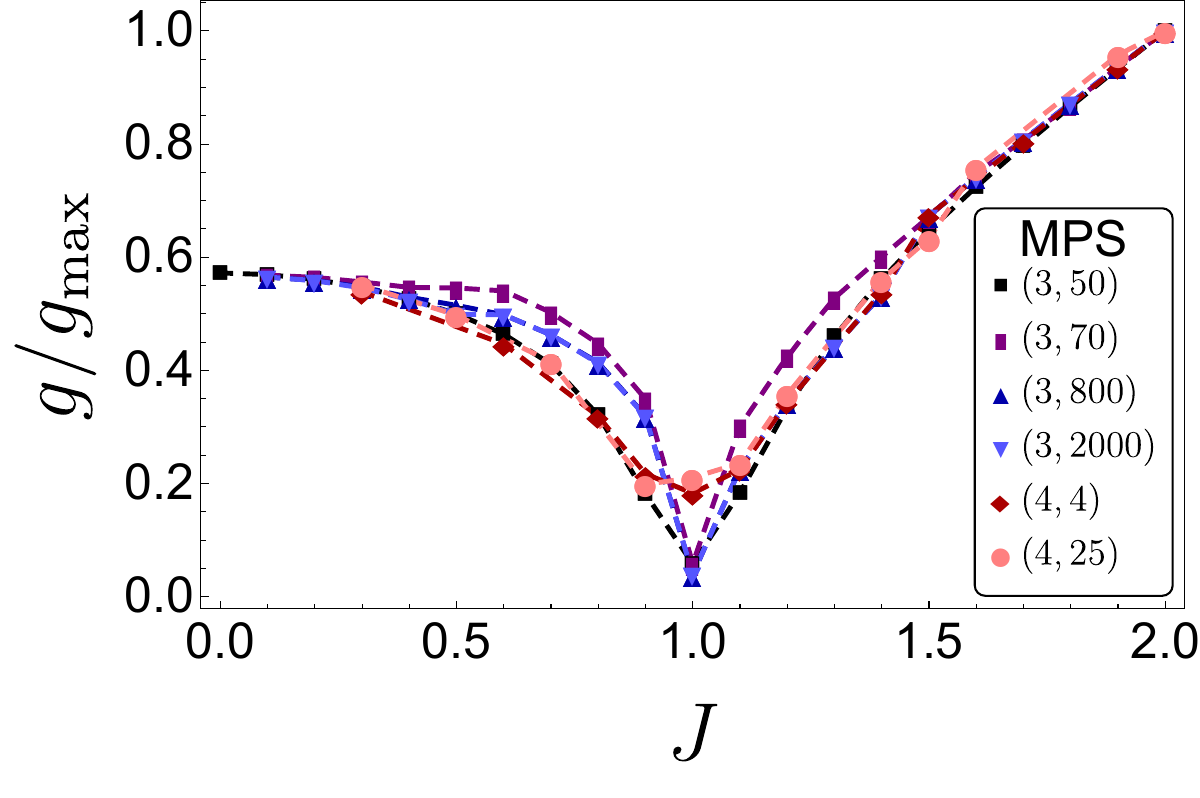}
    \end{subfigure}
    \begin{subfigure}[b]{0.45\textwidth}
        \includegraphics[width=8cm, height=5cm]{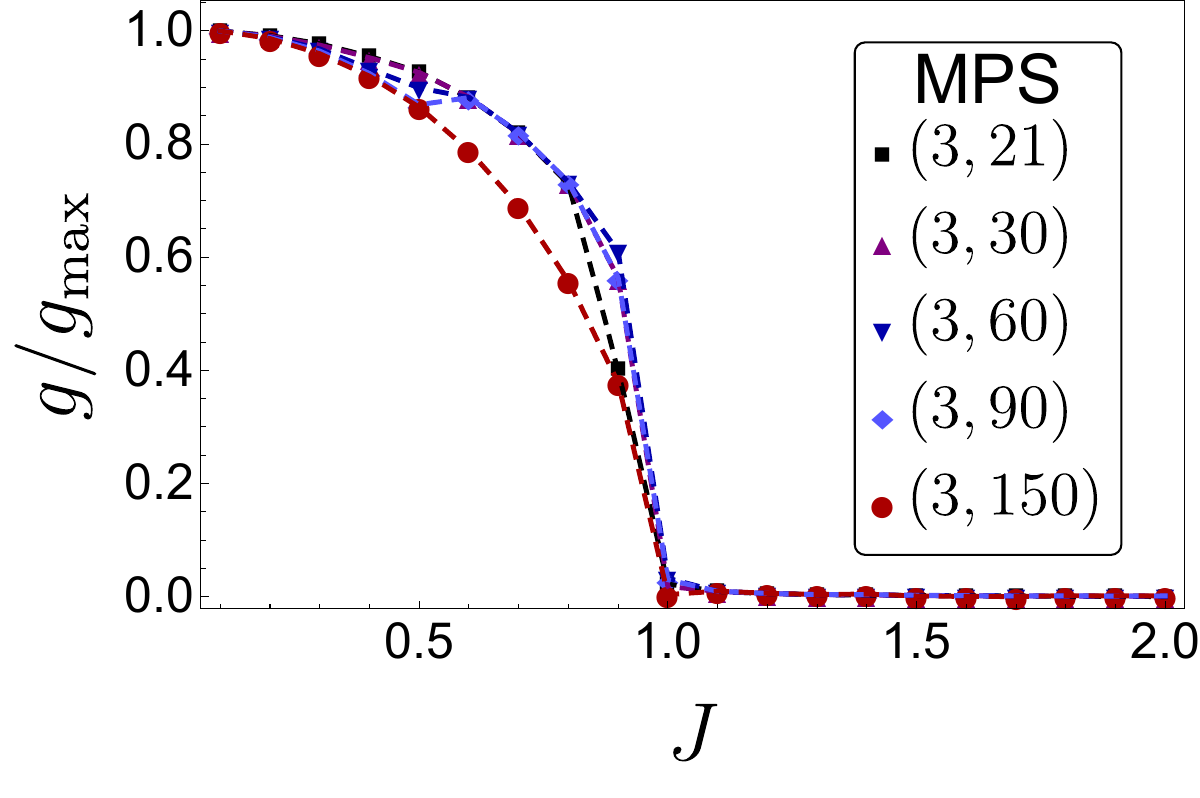}
    \end{subfigure}
    \caption{\label{fig:gap_MPS} 
        The gap, $g$, normalised with the maximum gap, $g_{\max}$, in the given domain for different lattice sizes from MPS without (a) and with (b) symmetries for the QTPM. For (a) $g_{\max} \approx 3.43-3.50$ and for (b) $g_{\max} = 2.0$.}
\end{figure*}

The same complications are encountered close to the phase transition from the MPS results in Fig.~\ref{fig:gap_MPS}(a). Both plots are normalised by the maximum value of the gap encountered in the region of $J$ values studied. For $0.0 < J <2.0$, the gap appears always to be maximum at $J = 2.0$ for Fig.~\ref{fig:gap_MPS}(a) and at $J = 0$ for Fig.~\ref{fig:gap_MPS}(b). In Fig.~\ref{fig:gap_MPS}(b), for $J>h = 1.0$ the gap approaches zero, as expected from the existence of degenerate ground states.

\end{document}